\documentclass{aa}

\usepackage{graphicx}
\usepackage{txfonts}
\usepackage{natbib}
\usepackage{threeparttable}
\usepackage{booktabs}
\usepackage{textgreek}
\usepackage{multirow}
\usepackage{orcidlink}
\usepackage[normalem]{ulem}
\usepackage{savefnmark}
\usepackage{mathtools}
\usepackage{array}
\usepackage{ragged2e}
\usepackage{soul}
\usepackage{float}

\newcommand{\Ha}{H\textalpha}
\newcommand{\Hb}{H\textbeta}

\newcommand{\Ion}[2]{#1\,{\sc #2}}
\newcommand{\Line}[3]{#1\,{\sc #2}~\textlambda #3}
\newcommand{\kms}{\mbox{$\mathrm{km~s^{-1}}$}}

\newcommand{\Msun}{\mbox{$M_{\rm \odot}$}}

\newcommand{\sw}{\textit{Swift}\,J1910.2--0546}
\newcommand{\jj}{J1910}

\newcommand{\ebv}{$E(B-V)$}
\newcommand{\ebvval}[1]{$E(B-V) = #1$}

\usepackage{hyperref}
\hypersetup{
    colorlinks=true,
    linkcolor=blue,
    citecolor=blue,
    urlcolor=blue
    }
    
\begin{document} 
\nolinenumbers
   \title{Characterising the short-orbital period X-ray transient \sw}

   \author{
          J.~M.~Corral-Santana
          \inst{1}
          \and
          P.~Rodríguez-Gil\inst{2,3}
          \and
          M.~A.~P.~Torres\inst{2,3}
          \and
          J.~Casares\inst{2,3}
          \and
          P.~G.~Jonker\inst{4,5}
          \and
          A.~Perdomo García\inst{6,2,3}
          \and
          D.~T.~Trelawny\inst{7}
          \and
          J.~A.~Carballo-Bello\inst{8}
          \and
          P.~A.~Charles\inst{9,10}
          \and
          D.~Mata Sánchez\inst{2,3}
          \and
          T.~Mu\~noz-Darias\inst{2,3}
          \and
          F.~A. Ringwald\inst{11}
          \and
          I.~G.~Martínez-Pais\inst{2,3}
          \and
          R.~L.~M.~Corradi\inst{12,2,3}
          \and
          P.~Saikia\inst{13}
          \and
          D.~M.~Russell\inst{13}
         }

   \institute{
             European Southern Observatory, Casilla 19001, Vitacura, Santiago, Chile \\
              \email{jcorral@eso.org}
         \and
             Instituto de Astrofísica de Canarias, E-38205 La Laguna, Tenerife, Spain  
         \and
             Departamento de Astrofísica, Universidad de La Laguna, E-38206 La Laguna, Tenerife, Spain  
         \and
            Department of Astrophysics/IMAPP, Radboud University Nijmegen, P.O. Box 9010, Nijmegen, 6500 GL, The Netherlands
         \and   
            SRON, Netherlands Institute for Space Research, Niels Bohrweg 4, Leiden, 2333 CA, The Netherlands
         \and
            Max Planck Institute for Astronomy, Königstuhl 17, 69117 Heidelberg, Germany
         \and
            Astronomy, Engineering, and Physics Department, Fresno City College, 1101 East University Ave., Fresno, CA 93741, USA
         \and   
            Instituto de Alta Investigación, Universidad de Tarapacá, Casilla 7D, Arica, Chile
         \and
            School of Physics \& Astronomy, University of Southampton, Southampton SO17 1BJ, UK
         \and
            Astrophysics, Department of Physics, University of Oxford, Keble Road, Oxford OX1 3RH, UK
         \and 
            Department of Physics, 2345 San Ramon Ave., M/S MH37, California State University, Fresno, CA 93740, USA
         \and
            Gran Telescopio Canarias, Cuesta de San José s/n, Breña Baja, 38712, Santa Cruz de Tenerife, Spain
         \and
            Center for Astro, Particle and Planetary Physics, New York University Abu Dhabi, PO Box 129188, Abu Dhabi, UAE  
         }

   \date{Received 16 July 2025; accepted 19 August 2025}

\abstract
{\sw\ (=MAXI~J1910$-$057) is a Galactic X-ray transient discovered during a bright outburst in 2012. Its X-ray spectral and timing properties point to a black-hole accretor, yet the orbital period remains uncertain, and no reliable dynamical constraints on the binary parameters are available.  
The 2012 event, extensively monitored at X-ray and optical wavelengths, offers a rare opportunity to investigate the structure and dynamics of the system and to constrain its fundamental properties.}
{We use time-series optical photometry and spectroscopy, obtained during outburst and quiescence, to estimate the orbital period, characterise the donor star, determine the interstellar extinction, distance, and system geometry, and constrain the component masses.}
{Multi-site $r$-band and clear-filter light curves and WHT/ACAM spectra from the 2012 outburst are combined with time-series spectroscopy from GTC/OSIRIS and VLT/FORS2 in quiescence.  Period searches are conducted using generalised Lomb–Scargle, phase-dispersion minimisation, and analysis-of-variance algorithms. Diffuse interstellar bands constrain $E(B\!-\!V)$, while empirical correlations involving \Ha\ yield estimates of $K_2$, $q$, and $i$.}
{We detect a coherent, double-humped modulation with a period of $0.0941\pm0.0007$\,d ($2.26\pm0.02$\,h) during the outburst. Its morphology is consistent with an early superhump, suggesting that the true orbital period may be slightly shorter than 4.52\,h. The \Ha\ radial velocity curves do not yield a definitive orbital period. In quiescence, TiO bands indicate an M3--M3.5 donor contributing $\simeq\!70\%$ of the red continuum. Diffuse interstellar bands give $E(B\!-\!V)=0.60\pm0.05$ and $N_{\mathrm{H}}=(3.9\pm1.3)\times10^{21}$\,cm$^{-2}$, placing the system at a distance of 2.8–4.0\,kpc. The \Ha\ line width in quiescence ($\mathrm{FWHM}_0 = 990\pm45$\,\kms), via a FWHM-$K_2$ calibration, provides an estimate of $K_2$, while its double-peaked profile gives $q$ and the orbital inclination. The latter appears much higher than estimates from X-ray studies. Adopting the resulting $K_{2}=230\pm17$\,\kms\ and $q=0.032\pm0.010$, and two orbital period scenarios (2.25 and 4.50\;h), Monte Carlo sampling returns a compact object mass $M_{1}=8$--$11$\,\Msun\ and an inclination $i = 13^{\circ}$–$18^{\circ}$ for plausible donor masses ($M_{2}=0.25$--$0.35$\,\Msun). We favour an orbital period of 4.5\;h.}
{\sw\ may be a short-period, low-inclination black hole X-ray transient, although a neutron star accretor cannot be completely ruled out. Further phase-resolved spectroscopy and photometry during quiescence are needed to better determine its fundamental parameters.} 


\authorrunning{Corral-Santana et al.}
\titlerunning{Characterising \sw}

\maketitle

\nolinenumbers

\section{Introduction}
\label{sec:intro}
Low-mass X-ray binaries (LMXBs) are systems in which either a stellar-mass
black hole (BH) or a neutron star accretes material from a low-mass companion star through
Roche-lobe overflow.
Among them, X-ray transient systems undergo sporadic outbursts driven by thermal-viscous instabilities in the accretion disc \citep{Lasota2001}, resulting in multi-wavelength brightness increases. 

During outburst, distinct accretion-related phenomena are observed in X-rays (e.g. \citealt{Done2007, Belloni2010}), while jets are observed in the radio band (e.g. \citealt{Fender2004}). The optical spectrum is typically dominated by emission lines formed in the outer disc atmosphere, which are sometimes observed to show signatures of accretion disc winds (e.g. \citealt{Munoz-Darias2019}). These are also detected in X-rays in highly inclined systems (e.g. \citealt{Ponti2012}).
As the system fades exponentially back to quiescence, rebrightening events may occur \citep[e.g.][]{Truss2002,Zurita2002,JimenezIbarra2019,Zhang2019}.

Thus far, BHs in Galactic LMXBs have been predominantly found in transient systems.
However, despite decades of study, only $\approx\!20$ out of 72 BH candidates have been dynamically confirmed (see BlackCAT\footnote{Online version available at \url{https://www.astro.puc.cl/BlackCAT} and \url{https://www.sc.eso.org/~jcorral/BlackCAT}}; \citealt{Corral-Santana2016}). This is due to the difficulty of performing dynamical studies during quiescence, where the donor star dominates the optical light \citep{Charles2006}. Moreover, constraining the BH mass requires measuring the orbital period and radial velocity amplitude of the donor, as well as additional parameters such as the binary mass ratio and system inclination.

\begin{figure}
    \label{fig:fc}
    \centering
	\includegraphics[width=0.95\columnwidth]{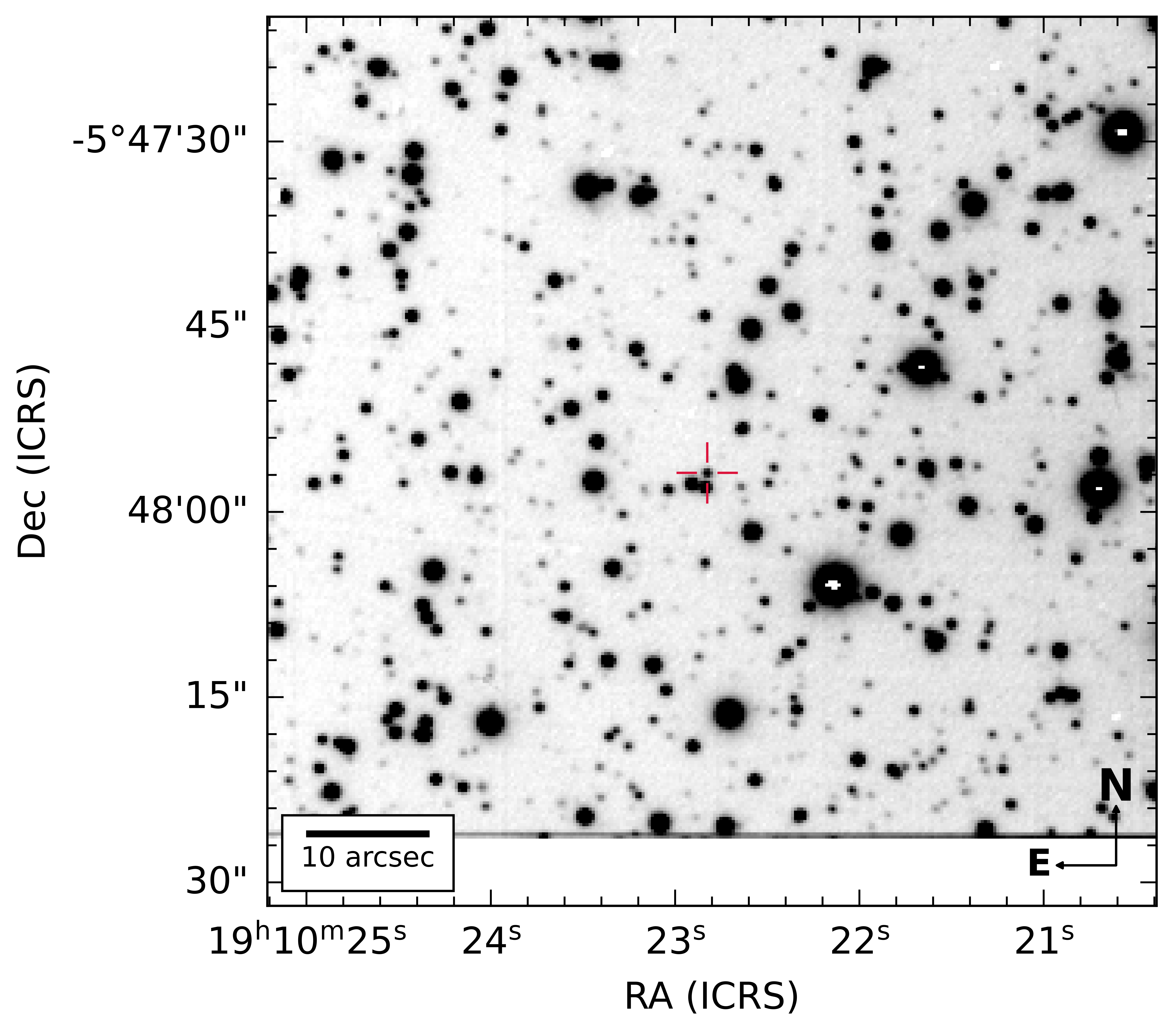}
    \caption{VLT/FORS2 $I$-band acquisition image obtained during quiescence. \jj\ is marked with a red cross at the centre of the 1.2\arcmin\ field. North is up, east is to the left. 
    }
\end{figure}

\sw\ (hereafter \jj), also known as MAXI~J1910$-$057, is a Galactic X-ray transient first detected on 30 May 2012 by the Burst Alert Telescope (BAT) on board the \textit{Neil Gehrels Swift Observatory}, and independently by the Monitor of All-sky X-ray Image/Gas Slit Camera (MAXI/GSC) \citep{Krimm2012, Usui2012}. \cite{Kimura2012} analysed a MAXI/GSC spectrum collected between 12 and 14 Jun 2012 and found that it is adequately modelled by a disc blackbody. The thermal nature of the emission is 
broadly consistent with the soft spectral state seen in other black hole candidates.
An optical/near-infrared counterpart was identified shortly after the onset of the outburst at R.A., Dec. (ICRS) = 19:10:22.79, –05:47:55.9 \citep[][see Fig.~\ref{fig:fc}]{Rau2012,Cenko2012}. 

Early optical spectroscopy taken on 18 Jun 2012 revealed an almost featureless blue continuum, lacking Balmer or \Ion{He}{i} lines, but displaying a broad \Line{He}{ii}{4686} emission line \citep{Charles2012}. Subsequent spectroscopy, revisited here (see Sect.~\ref{sec:obs-out} for details), showed a weak \Line{He}{ii}{4686} emission line, broad H$\beta$ and H$\gamma$ absorptions, and a wide \Ha\ absorption trough with a weak emission component as reported in \cite{Casares2012}, who tentatively constrained the orbital period to be $> 6.2$\;h. However, \citet{Lloyd2012} had earlier reported a putative modulation at 2.25--2.47\,h, which, if confirmed, would make \jj~the shortest orbital period BH X-ray binary known to date. 

By early 2013, the system was considered in quiescence \citep{Nakahira2014,Saikia2023}. 
In Jul 2015, \cite{Lopez2019} obtained optical/near-infrared photometry, reporting AB magnitudes of $r = 23.46 \pm 0.07$, $i = 22.18 \pm 0.04$, and $K_\mathrm{s} = 20.43 \pm 0.11$. However, \cite{Saikia2023} later reported values approximately one mag brighter around the same time (see their fig.~8), attributing this discrepancy to photometry contamination from nearby field stars or to intrinsic accretion variability within the system.

\jj~remained in quiescence for about ten years, showing occasional reflares and rebrightenings during that period \citep[see][for a comprehensive study]{Saikia2023} until 8 Feb 2022, when \cite{Tominaga2022} reported a new outburst  detected by MAXI. This event was also observed in radio \citep{Williams2022} and optical wavelengths \citep{Hosokawa2022,Kong2022}. By late Mar 2022, the system was reported to have returned to quiescence \citep{Saikia2022}.\\

In this paper, we present the analysis of time-series optical photometry and spectroscopy of \jj\ during its 2012 outburst and subsequent decline, along with additional spectroscopy obtained in 2016 when the system was in quiescence. Section \ref{sec:obs} details the observations and data reduction processes. Section~\ref{sec:res} presents the results, including an estimate of the orbital period,
spectral type of the donor star, estimation of the distance, and constraints on the binary parameters.
Section~\ref{sec:discuss} discusses the implications of our findings, and Section~\ref{sec:con} contains our conclusions.

\section{Observations and data reduction}
\label{sec:obs}
\subsection{Optical photometry and spectroscopy during outburst}
\label{sec:obs-out}

Time-series photometry of \jj\ in the $r$-band was obtained in Jun 2012 (see Table~\ref{tab:phot} for a detailed log of the observations) using the Wide Field Camera (WFC)
mounted on the 2.5-m Isaac Newton Telescope (INT)
at the Roque de los Muchachos Observatory (La Palma). The WFC is a prime-focus mosaic of four $2048 \times 4100$ pixel EEV42 CCDs, offering a field of view of approximately $34\arcmin \times 34\arcmin$. In this study, only CCD \#4 was used for imaging with an exposure time of 5\;s.
We conducted variable-aperture photometry of \jj\ relative to the comparison star \textit{Gaia} DR3 4206052097085749760 \citep[$r = 15.311 \pm 0.001$ in Pan-STARRS1;][]{Chambers2016} using the HiPERCAM reduction software\footnote{\url{https://github.com/HiPERCAM/hipercam}}.

\begin{table}
    \centering
    \caption{Log of photometric data obtained in 2012.}
	\label{tab:phot}
    \begin{threeparttable}
    \begin{tabular}{lccccc} 
     \toprule\noalign{\smallskip}
   Date      & Filter      & Exp.     & No.       & Time     & Telescope/               \\
             &             & time     & of        & span     & Instrument$^\mathrm{a}$  \\
             &             & (s)      & images    & (h)      &                          \\
\midrule\noalign{\smallskip}
04 Jun       & $r$         & 120      & 113       & 6.52     & SMARTS 0.9-m             \\
09 Jun       & $r$         & 5        & 938       & 3.74     & INT/WFC                  \\
10 Jun       & $r$         & 5        & 1106      & 4.85     & INT/WFC                  \\
11 Jun       & $r$         & 5        & 1124      & 4.05     & INT/WFC                  \\
14 Jun       & $r$         & 5        & 810       & 2.82     & INT/WFC                  \\
16 Jun       & $r$         & 5        & 634       & 4.40     & INT/WFC                  \\
27 Jun       & $C$$^\mathrm{b}$       & 120      & 95        & 3.75     & SRO 0.41-m        \\
28 Jun       & $C$       & 120      & 166       & 6.29     & SRO 0.41-m        \\
10 Jul       & $C$       & 120      & 180       & 6.75     & SRO 0.41-m        \\
20 Jul       & $C$       & 120      & 147       & 5.52     & SRO 0.41-m        \\
        \bottomrule\noalign{\smallskip}
\end{tabular}
\begin{tablenotes}\footnotesize
\item[a] WFC: Wide Field Camera.
\item[b] Clear filter.
\end{tablenotes}
\end{threeparttable}
\end{table}

We extended the time baseline by adding time-series $r$-band imaging with the Small and Moderate Aperture Research Telescope System (SMARTS) Consortium 0.9-m telescope on Cerro Tololo (Chile), conducted on the night of 4 Jun 2012\footnote{We downloaded the SMARTS 0.9-m images from \url{https://astroarchive.noirlab.edu/portal/search}}, as reported in \cite{Britt2012}. The $2048 \times 2048$\,pixel Tek2K CCD detector was used with exposure times of 120\,s. We performed relative photometry on the reduced images using the HiPERCAM pipeline and the same comparison star as that used for the WFC photometry. 

\begin{figure*}
\sidecaption
  \includegraphics[width=12cm]{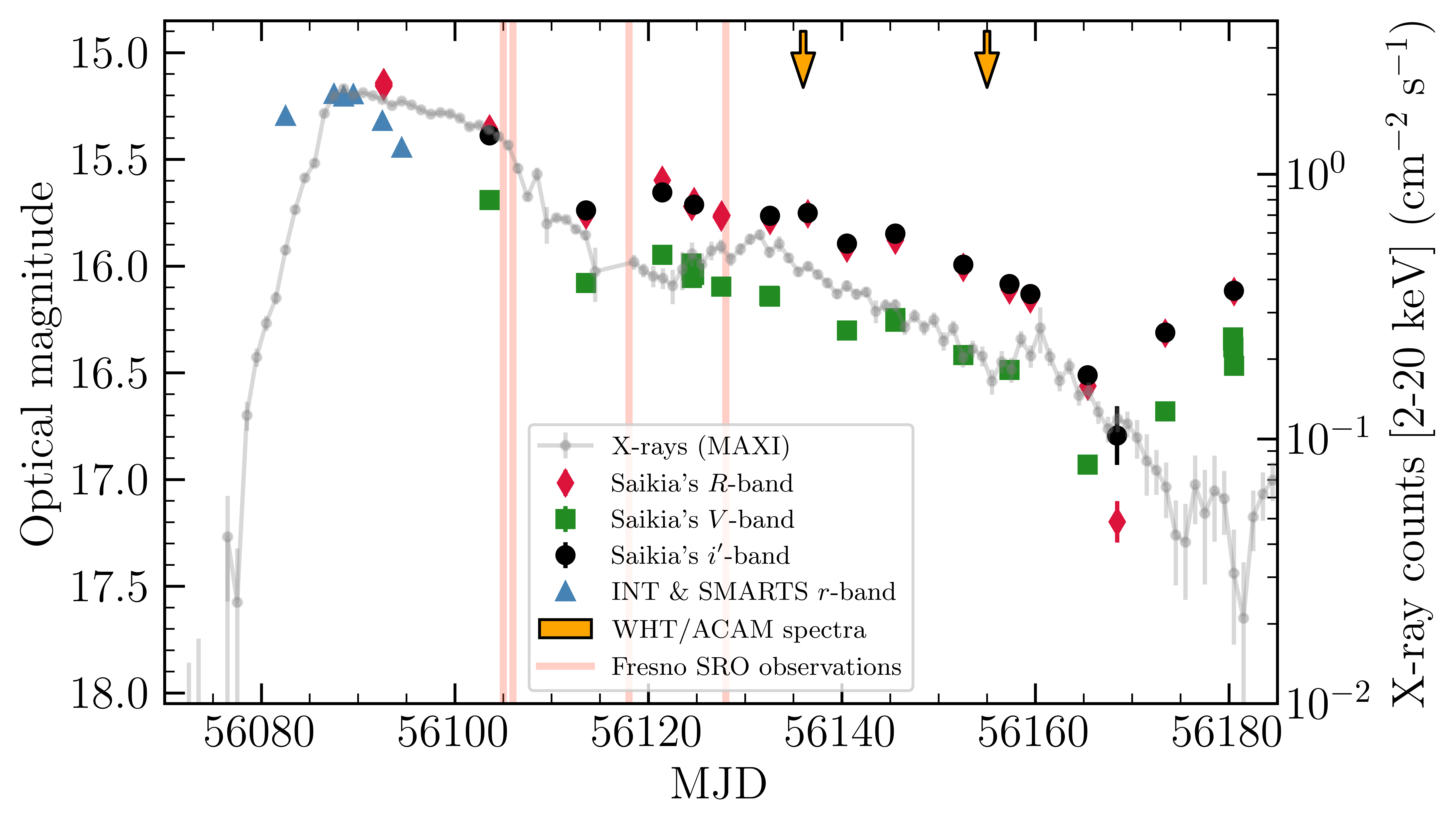}
     \caption{Evolution of the 2012 outburst. X-ray data are shown as a grey line, while the $R$-, $V$-, and $i'$-band magnitudes published by \cite{Saikia2023} are represented by red diamonds, green squares, and black circles, respectively. Time-series photometry obtained with the SMARTS 0.9-m and INT/WFC in the $r$-band is marked with blue triangles, whereas the time of the WHT/ACAM spectra are indicated by orange arrows. Clear-filter photometry from Fresno SRO is marked by vertical red lines.\\
    }
     \label{fig:outevol}
\end{figure*}

In Sect.~\ref{sec:Psearch}, we also reanalyse some of the 2012 photometric data sets originally presented in the Master's thesis of co-author Trelawny \citep{Trelawny2013}. Specifically, we focus on the data collected during 27, 28 Jun and 10, 20 Jul 2012, which exhibited the clearest variability. The images were obtained with the DFM Engineering 0.41-m $f$/8 telescope at Fresno State’s facility of the Sierra Remote Observatories (SRO), equipped with a $4008 \times 2672$ pixel Santa Barbara Instruments Group STL--11000M CCD camera binned $3 \times 3$, and an Astrodon clear filter. The Clear filter had a nearly uniform, 95--98\% transparency to light with wavelengths of 3750--12000\;\AA. Each frame had an exposure time of 120\,s. All data were processed using the \verb|AIP4WIN| 2.1.8 software \citep{Berry2005}. 

We also obtained time-series optical spectroscopy of \jj\ on 28 Jul and 16 Aug 2012, a few weeks after the beginning of the outburst, using the Auxiliary-port CAMera (ACAM) mounted on the 4.2-m William Herschel Telescope (WHT) on La Palma. We obtained thirty-five 300-s (signal-to-noise ratio, S/N, of approximately~$40\!-\!60$ per pixel), and ten 600-s ($\mathrm{S/N}\!\approx\!40\!-\!75$ per pixel) spectra, respectively, using a 0.5\arcsec\ slit width. The 400-line/mm$^{-1}$ transmission Volume Phase Holographic grating (\textlambda\textlambda3950--9400) was used,
with a resolving power $\lambda/\Delta \lambda\!\simeq\! 900$. The full width at half maximum (FWHM) of the night-sky [\ion{O}{i}] emission line at 5577.338\,\AA\ was measured to be $\simeq\! 6.3$\,\AA\ ($\simeq\!290$~\kms), taken as $\Delta\lambda$.
The average seeing was approximately 0.7\arcsec\ on both nights, so the spectra were slit-limited. The data were de-biased, flat-fielded and wavelength-calibrated using standard procedures, and cover the range \textlambda\textlambda$6000-8000$. The rms scatter of the pixel-to-wavelength calibration fit is $\simeq 0.1$\;\AA.

Fig.~\ref{fig:outevol} shows the evolution of the 2012 outburst in X-rays and in three of the optical bands of \cite{Saikia2023}. In the figure, we mark the dates on which the WHT/ACAM time-resolved spectroscopy and the photometry mentioned above and listed in Table~\ref{tab:phot} were obtained.

\subsection{Optical spectroscopy and photometry during quiescence}
\label{sec:obs-qui}
Five 1800-s spectra were obtained on 1 Aug 2016, approximately 1500 days after discovery, when \jj\ was about eight magnitudes fainter than at the peak of the outburst. 

For these observations, we used the Optical System for Imaging and low-Intermediate-Resolution Integrated Spectroscopy (OSIRIS) instrument \citep{Cepa2003} mounted on the 10.4-m Gran Telescopio Canarias (GTC) at the Roque de los Muchachos Observatory. The data were taken using the R1000R grism (\textlambda\textlambda5200--10200, mean reciprocal dispersion of 2.6\;\AA\;pixel$^{-1}$) with a 0.8\arcsec\ slit width, providing a resolution of approximately 7.1\;\AA, and a S/N per pixel of $\approx\!10$ in the continuum close to \Ha. The rms scatter of the pixel-to-wavelength calibration fit is 0.11\;\AA. The average seeing conditions during that night were around 1\arcsec. The reduction process for these spectra was similar to that used for the outburst WHT/ACAM data. 
To correct for the instrumental response, we used the spectrophotometric standard Feige~66 \citep{Oke1990}.

Note that the OSIRIS $i$-band acquisition images show that \jj's PSF is contaminated by two much brighter stars located just to the south. Although these sources are well resolved in Fig.~\ref{fig:fc}, the relatively poor seeing conditions during the OSIRIS observations allowed flux from the nearby stars to enter the slit. This resulted in a dilution of the target spectrum, which may contribute to the differences discussed in  
Section~\ref{sec:Ha_RVC_above_quies}.

Further optical spectroscopy was conducted using the Very Large Telescope (VLT) located at Cerro Paranal, Chile, while \jj\ was in quiescence. The FORS2 spectrograph was configured with the 600RI grism and a 1\arcsec\ slit width, yielding a useful spectral range of 5330--8385\;\AA. The average seeing during the observations, as measured by collapsing the 2D spectra along the dispersion axis in the vicinity of \Ha, ranged from 0.35\arcsec\ to 0.94\arcsec, resulting in a spectral resolution of 3.3--5.0\;\AA, corresponding to velocity resolutions of approximately 160--240\,\kms~at 6300\,\AA.

The spectra were acquired with the MIT/LL mosaic detector, binned $2\!\times\!2$. The observations took place between 30 Apr and 5 Sep 2016, during which a total of 15 spectra were obtained. Each exposure had a duration of 1270\;s, and the S/N per pixel of the extracted spectra in the vicinity of \Ha\ is in the range $\approx\!2-\!8$. As mentioned earlier, the average seeing during these exposures ranged from 0.35\arcsec\ to 0.94\arcsec, providing sufficient spatial resolution to distinguish \jj\ from the nearby brighter star located 1.3\arcsec\ south of the target (see Fig.~\ref{fig:fc}).
As this star was simultaneously included on the slit, we used its spectrum to correct for telluric absorptions after masking out any stellar features.
We corrected the FORS2 spectra of \jj\ for instrumental response using an ESO archive spectrum of the standard star LTT\;7379 \citep{Hamuy1992, Hamuy1994}, obtained on the night of 3 Sep 2016.

The acquisition images taken prior to each FORS2 spectrum were obtained in the $I$-band using the I\_BESS+77 filter, with an exposure time of 120\,s. These images were processed using point spread function (PSF) photometry with \texttt{DAOPHOT II/ALLSTAR} \citep{Stetson1987}.
To calibrate our $I$-band photometry, we cross-matched the resulting source catalogue with Pan-STARRS1 \citep[AB magnitudes;][]{Chambers2016} and converted these magnitudes to the Johnson-Cousins $I$-band using the transformation equations provided in table~6 of \citet{Tonry2012}. Approximately 50 stellar-like sources per image (defined by a sharpness value $\leq 0.1$) with relatively low PSF photometric uncertainties ($\leq 0.05$\,mag) were used to derive the individual zero-points. The resulting magnitudes are listed in  Table~\ref{tab:FORS2_quies}.

\begin{table}
	\centering
     		\caption{Log of 2016 VLT/FORS2 spectroscopic observations of \jj\ during quiescence.
            }
	\label{tab:FORS2_quies}
    \begin{threeparttable}
    \begin{tabular}{lcc}
     \toprule\noalign{\smallskip}
2016 UT date          &  Seeing    & Acq. image \\
at start         & (\arcsec)  & $I$-mag (Vega)   \\

\midrule\noalign{\smallskip}
01.37 May   & 0.50  & $21.54\pm0.05$ \\  
02.27 May   & 0.58  & $21.55\pm0.04$ \\ 
02.34 May   & 0.73  & $21.53\pm0.03$ \\  
11.33 Jun   & 0.55  & $21.62\pm0.04$ \\ 
01.06 Jul   & 0.83  & $21.55\pm0.02$ \\ 
01.11 Jul   & 0.63  & $21.57\pm0.04$ \\  
01.14 Jul   & 0.50  & $21.55\pm0.04$ \\  
01.22 Jul   & 0.61  & $21.66\pm0.04$ \\  
02.09 Jul   & 0.67  & $21.58\pm0.03$ \\ 
02.12 Jul   & 0.70  & $21.62\pm0.04$ \\ 
07.23 Jul   & 0.79  & $21.69\pm0.02$ \\ 
04.09 Sep   & 0.52  & $21.63\pm0.06$ \\ 
04.12 Sep   & 0.54  & $21.69\pm0.06$ \\ 
05.04 Sep   & 0.70  & $21.52\pm0.03$ \\ 
06.03 Sep   & 0.62  & $21.59\pm0.04$ \\
\bottomrule\noalign{\smallskip}
\end{tabular}
\end{threeparttable}
\end{table}

We also performed PSF photometry on the average OSIRIS $i$-band acquisition image using \texttt{DAOPHOT\,II/ALLSTAR}, and estimated the zero-point by following a procedure similar to that used for the FORS2 acquisition images.
With a total of 37 stars in common with Pan-STARRS1 in our catalogue, we derived $i = 22.18\pm0.03$ for \jj, identical to that reported in \cite{Lopez2019}. A comparison with the FORS2 PSF catalogues indicates a systematic difference of approximately 0.54\;mag, which accounts for the different photometric systems. This would correspond to an OSIRIS magnitude of roughly $I \simeq 21.65$, consistent with the values in Table~\ref{tab:FORS2_quies}.

\section{Results}
\label{sec:res}

\begin{figure*}
	\includegraphics[width=0.9\textwidth]{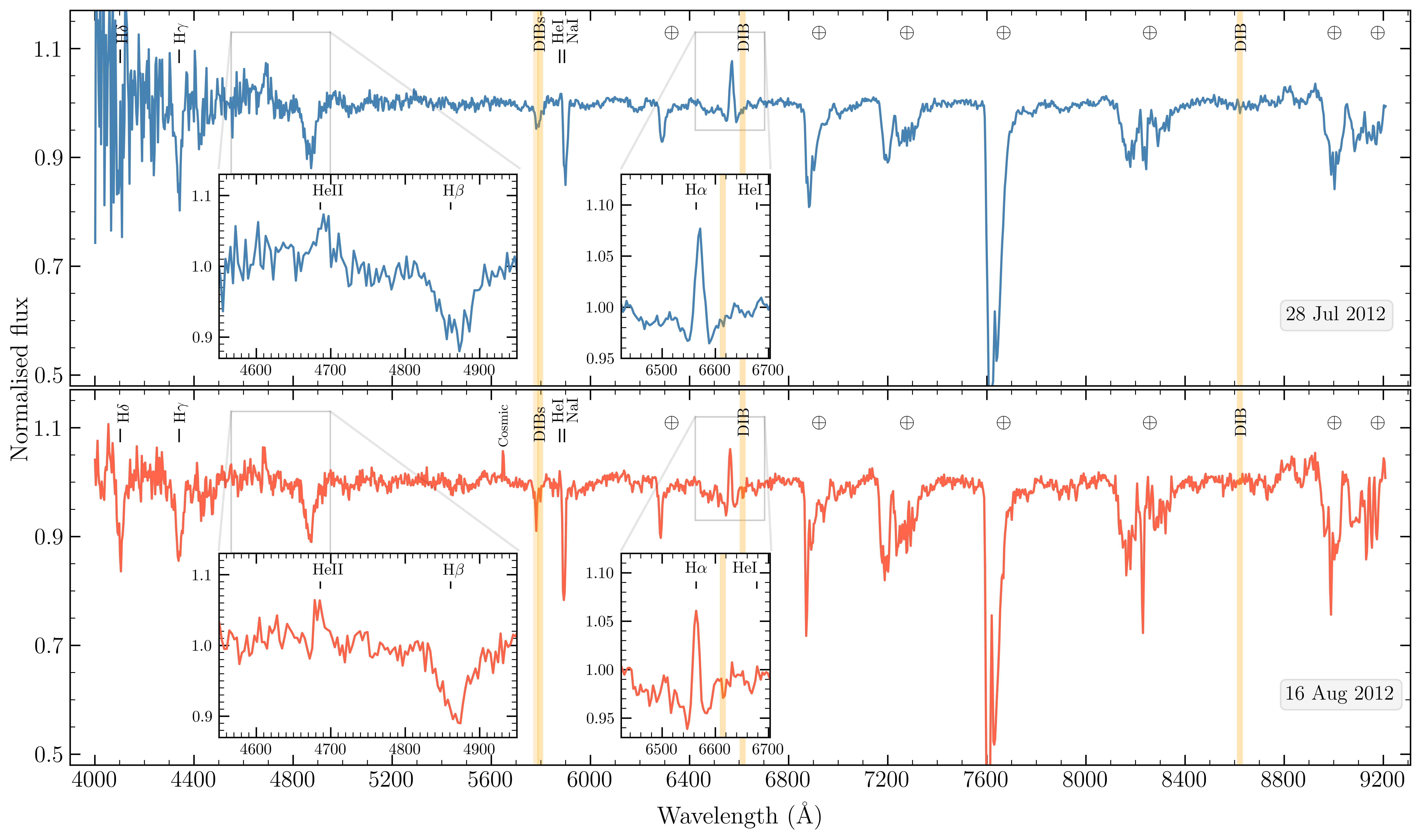}
    \caption{Average spectra of \jj\ during outburst from the first (upper; blue) and
second (lower; red) nights, obtained with ACAM after continuum
normalisation. The Balmer series down to H$\delta$ is visible, exhibiting
both emission and absorption components. Zoomed-in regions show the areas
around \Ha\ and \Hb. \Ion{He}{i} and \Ion{He}{ii} lines are also present
and identified in the spectra, along with several diffuse interstellar
bands marked with orange bars. 
    }
    \label{fig:spec}
\end{figure*}

\subsection{The outburst spectrum}
\label{sec:incl}
During the 2012 outburst, the ACAM spectra of \jj\ exhibit a range of spectral features, as shown in Fig.~\ref{fig:spec}. The Balmer series is detected, from \Ha\ to H$\delta$, with the lines displaying both emission and broader absorption components. \Ha\ is the strongest of the emission lines, characterised by a narrow, single-peaked profile. Weaker emission lines of \Line{He}{i}{5876} and possibly \textlambda$6678$ are also present, while the \Line{He}{ii}{4686} emission line is clearly detected.

We measured the intrinsic \Ha\ FWHM (FWHM$_0$, corrected for the instrumental profile) 
by fitting Gaussian profiles, obtaining values of $540\pm20$ and $570\pm40$\;\kms~for the first and second night, respectively. The corresponding equivalent widths (EW) are $2.30\pm0.03$ and $1.83\pm0.04$\,\AA. 
Although there are no significant differences in the average spectra of both nights (see Fig.~\ref{fig:spec}), we chose not to combine them, as \jj\ was 
in a different X-ray state on each occasion, with the soft X-ray flux rapidly decreasing between the two epochs (see Fig.~\ref{fig:outevol}; see also \citealt{Nakahira2012,Nakahira2014}). 
The measured FWHM values are consistent with those observed in low-inclination systems in outburst, such as MAXI\,J1836$-$194 \citep[$\approx 270$\;\kms;][]{Russell2014}. 

Interstellar absorption features are also present, the most prominent being the unresolved \Ion{Na}{i} doublet adjacent to \Line{He}{i}{5876}. Several diffuse interstellar bands (DIBs) are detected as well; three of them, at 5780, 5797, and 6614\,\AA, will be later used to estimate the colour excess (Sect.~\ref{sec:redd}).

   \begin{figure*}
   \centering
   \includegraphics[width=0.95\textwidth]{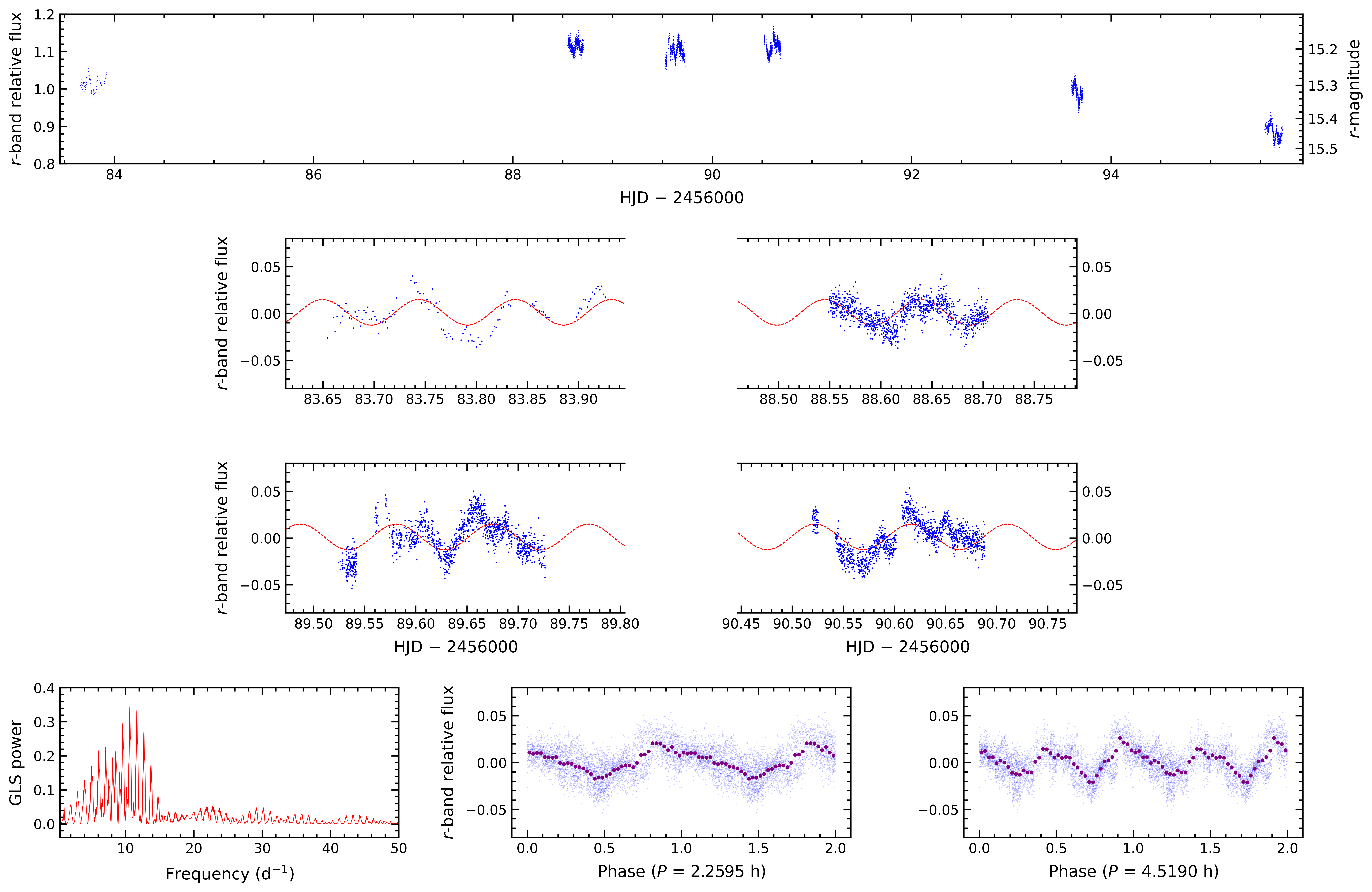}
    \caption{Top panel: $r$-band light curves from SMARTS 0.9-m and INT/WFC, expressed as flux ratios and magnitudes, relative to the comparison star. Middle panels: SMARTS light curve (top left) and INT/WFC light curves from the first three nights when \jj\ maintained an almost constant flux level, after subtracting the nightly averages. The red dashed curve represents the best-fit sine wave with a period corresponding to the highest peak in the periodogram shown below. Bottom left panel: GLS periodogram of the four $r$-band light curves displayed in the middle panels (4, 9, 10, and 11 Jun 2012). The highest peak corresponds to a period of $0.0941$\;d ($= 2.26$\;h). Bottom middle and right panels: The four light curves folded on this period and twice that value. The pale blue points represent the unbinned data, while the purple points are binned data across 40 phase intervals. Zero phase corresponds to the HJD of the first data point, and a full cycle is repeated for continuity.} 
	\label{j1910_mnras_fig_GLS_01}
    \end{figure*}

   \begin{figure*}
   \centering
   \includegraphics[width=0.95\textwidth]{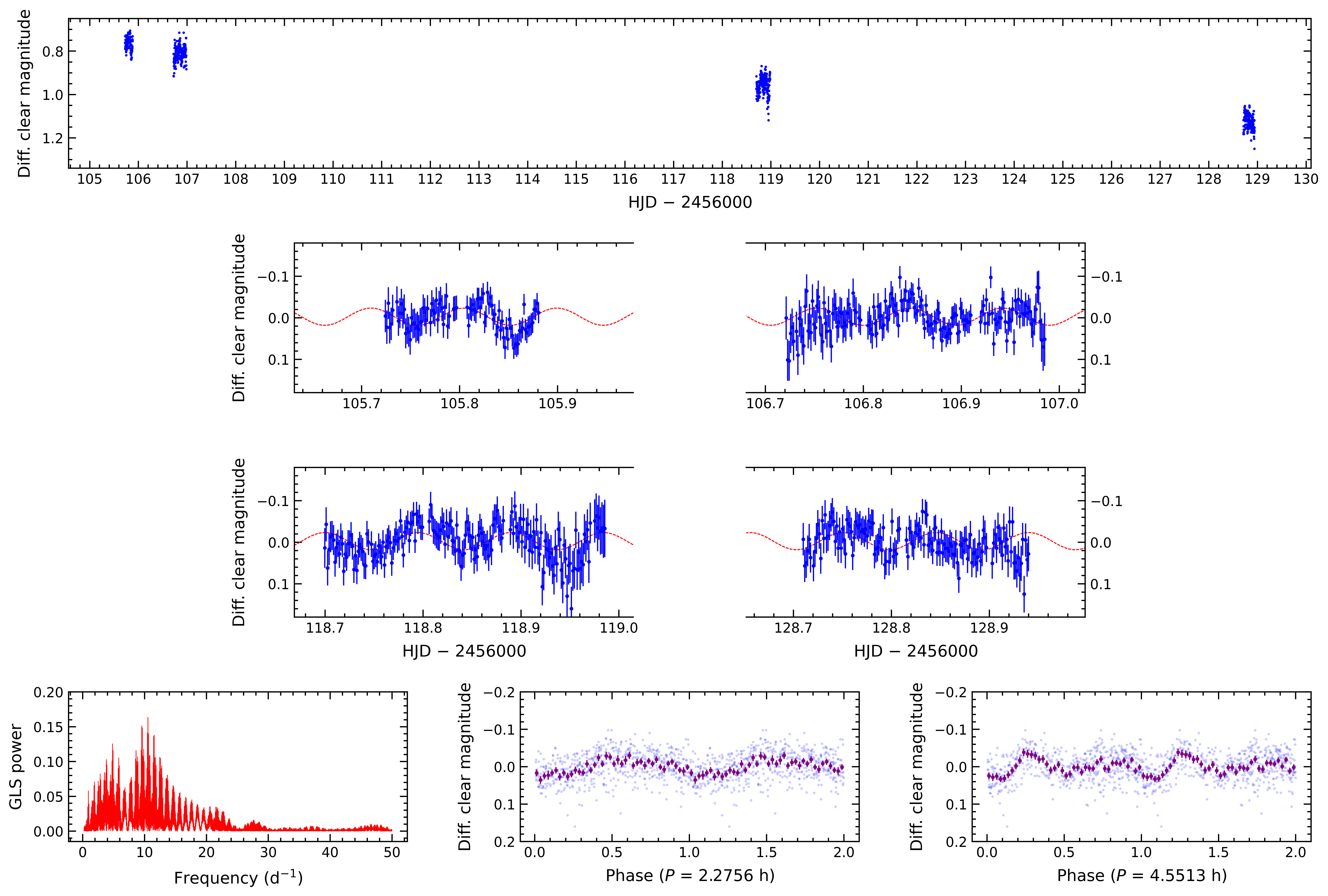}
    \caption{Top panel: Clear-filter light curves from Fresno SRO on 27, 28 Jun, and 10, 20 Jul 2012. Middle panels: Zoomed-in view of the four light curves. The red dashed curve represents the best-fit sine wave with a period corresponding to the highest peak in the periodogram shown below. Bottom left panel: GLS periodogram of the four clear-filter light curves shown in the middle panels after subtraction of nightly averages. The highest peak corresponds to a period of $0.0948$\;d ($2.275$\;h). Bottom middle and right panels: The four light curves folded on this period and twice that value. The pale blue points represent the unbinned data, while the purple points are binned data across 40 phase intervals. Zero phase is the HJD of the first data point.} 
	\label{fig:j1910_mnras_fig_GLS_june_july_2012}
    \end{figure*}

\subsection{Light curve period search}\label{sec:Psearch}
We conducted a search for periodicities in the SMARTS $r$-band light curve obtained on 4 Jun 2012, about six days 
before the peak of the 2--20\;keV X-ray light curve (see Fig.~\ref{fig:outevol}), together with the WFC $r$-band light curves from 9, 10, and 11 Jun 2012, during the outburst phase (corresponding to the first four light curves in Fig.~\ref{j1910_mnras_fig_GLS_01}). This  analysis specifically targets epochs where the system maintained an almost constant average flux level, with a nightly time baseline exceeding 3.5 hours. 

Prior to the period search, we subtracted the nightly mean relative flux values. The resulting light curves, depicted in the middle panels of Fig.~\ref{j1910_mnras_fig_GLS_01}, underwent a generalised Lomb-Scargle analysis \citep[GLS;][]{Zechmeister2009}, based on a fitting function that includes a constant term and a sine wave. We employed the GLS class from \textsc{PyAstronomy}\footnote{\url{https://github.com/sczesla/PyAstronomy}} \citep{Czesla2019} for this analysis. 

We first estimated the period uncertainty using a bootstrap approach with flux perturbation. For each of 100\,000 iterations, the light curve was resampled with replacement and the fluxes were randomly perturbed within their uncertainties. A GLS periodogram was computed for each realisation, and the frequency of the highest peak was recorded. The resulting distribution of peak periods was fitted with a Gaussian over a restricted interval around the dominant peak, and the mean and standard deviation were adopted as the representative period and its 1$\sigma$ uncertainty. While this bootstrap-based method provides a formal estimate of the period uncertainty, we find that it can yield unrealistically small values, typically on the order of a few $10^{-5}$\;d. Such precision is inconsistent with the width of the corresponding peaks in the periodograms, which often have full widths at half maximum $\mathrm{FWHM}\!\sim\!0.1$\;d$^{-1}$. This discrepancy indicates that the bootstrap approach underestimates the true uncertainty, likely because it captures only the statistical noise in the flux values and not broader effects such as spectral leakage, aliasing, or the finite frequency resolution imposed by the time baseline. To obtain a more realistic estimate, we adopt as the period uncertainty the value implied by the FWHM of the main periodogram peak, converted to the period domain.

The bottom-left panel of Fig.~\ref{j1910_mnras_fig_GLS_01} illustrates the resulting periodogram, which yields a preferred period of $0.0941(5)$\,d ($=\!2.26(1)$\,h; the numbers in parentheses indicate the $1\sigma$ uncertainties in the last digits). Consistent results were obtained using alternative period search methods, including phase dispersion minimisation \citep[PDM;][]{Stellingwerf1978} and analysis of variance \citep[AOV;][]{Schwarzenberg-Czerny1996} techniques.
The bottom-middle panel shows the $r$-band light curves from 4, 9, 10, and 11 Jun 2012, folded on the detected period (pale blue dots) and binned into 40 phase intervals (purple dots). This period is 
close to the values of $0.09353(2)$\,d ($=2.2447(5)$\,h) and $0.09245(2)$\,d ($=2.2188(5)$\,h) reported by \cite{Lloyd2012}, which were derived from observations conducted between 14 Jun and 7 Jul 2012, with coverage of 2 to 7\,hours per night. 

We repeated the analysis using all the light curves, subtracting the nightly linear trends of the last two to account for the overall brightness decline. The highest peak in the GLS periodogram also corresponds to a period of 2.26(1)\;h. 

We also subjected the Jun and Jul 2012 Fresno SRO light curves to a GLS period search (illustrated in Fig.~\ref{fig:j1910_mnras_fig_GLS_june_july_2012}), similar to the analysis performed on the combined SMARTS and WFC time-series data. The preferred period is $0.0948(1)$\,d ($=2.275(3)$\,h). In this case, the AOV periodogram shows the strongest peak near 0.19\;d, while the PDM periodogram displays a corresponding minimum at the same period; both also contain signal near half that value at approximately 0.095\;d.

Finally, a GLS period analysis of the combined light curve set (SMARTS, WFC, and Fresno SRO) yields a best period of $0.0941(7)$\,d ($=2.26(2)$\,h). 
A summary of the light curve period search results is provided in Table~\ref{tab:phot_periods}.

\cite{Saikia2023a} tentatively suggested an orbital period in the range 2.25--2.47\,h based on their analysis of $B$-band photometry obtained on 16 and 19 Jul 2012. Upon re-examining their data, we find that only the observations from the first night (represented by red circles in their fig.~4) may exhibit a distinct modulation. Our sine-wave fit to these data yields a periodicity of $0.103(1)$\,d ($= 2.47(2)$\,h). However, the short duration and sparse sampling of the original observations limit the accuracy of any period determination, especially in comparison to the more extensive dataset presented here. 

\begin{table*}
	\centering
     		\caption{Results of the light curve period search.}
	\label{tab:phot_periods}
    \begin{threeparttable}
	\begin{tabular}{lccc} 
     \toprule\noalign{\smallskip}
		Data set (2012 date) & Filter & Flux trend & GLS best period \\
                 &        &         & (h)\\
 		\midrule\noalign{\smallskip}
		SMARTS (4 Jun) + WFC (9--11 Jun) & $r$ & Constant & $2.26(1)$ \\
       \noalign{\smallskip}
	       \noalign{\smallskip}
	   Fresno SRO (27, 28 Jun; 10, 20 Jul) & Clear & Decline & $2.275(3)$ \\
	       \noalign{\smallskip}
      All combined & $r$, Clear &  & $2.26(2)$ \\
       \noalign{\smallskip}
        \bottomrule\noalign{\smallskip}
\end{tabular}
\end{threeparttable}
\end{table*}

\begin{figure}
	\includegraphics[width=0.97\columnwidth]{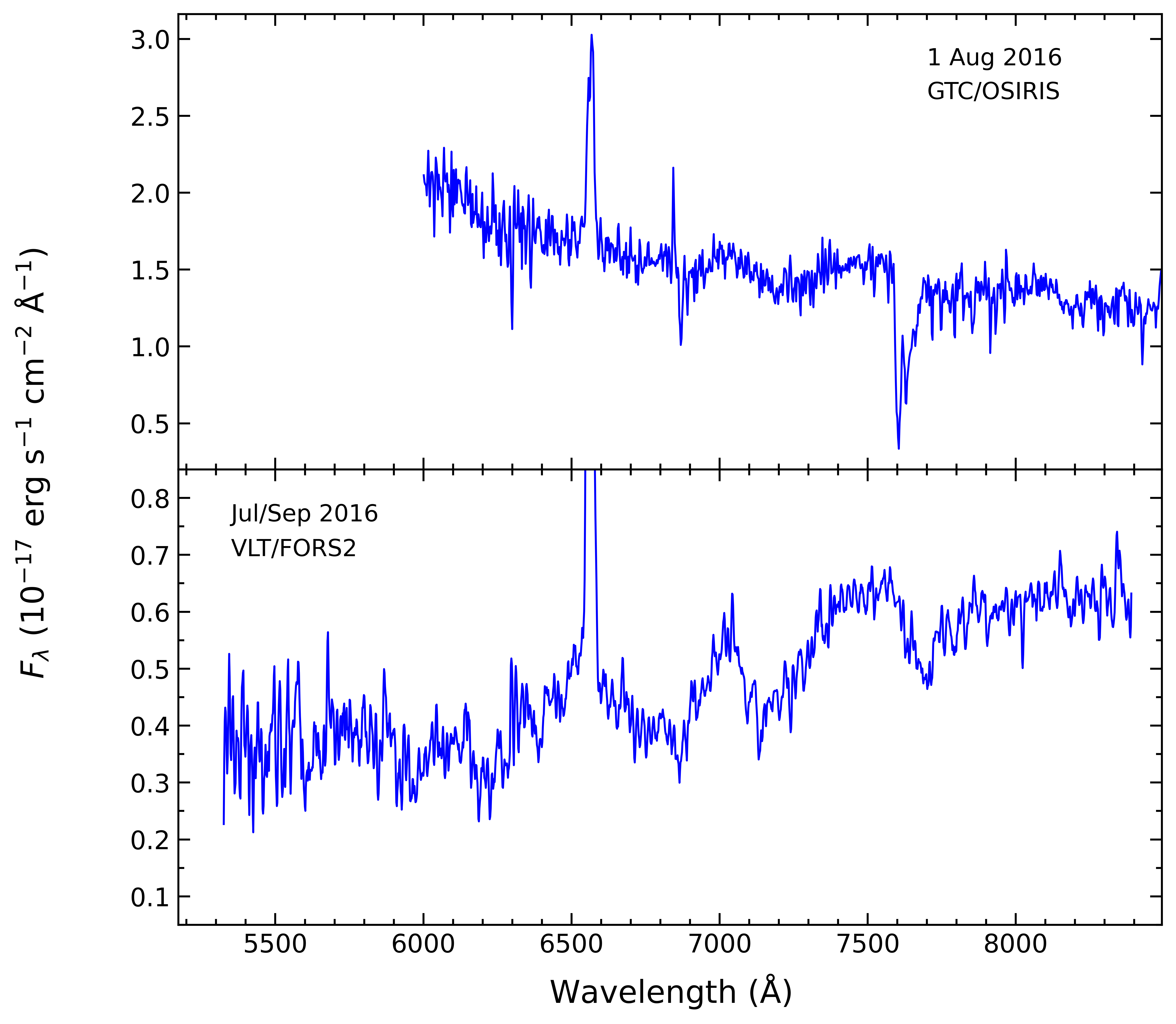}
    \caption{Top panel: Average GTC/OSIRIS spectrum of \jj\ during quiescence. The bluer part has been trimmed due to unreliable correction for the instrumental response. Although veiled by slit contamination from the two nearby stars (see Fig.~\ref{fig:fc}), caused by seeing $\gtrsim\!1$\arcsec, 
    TiO absorption bands characteristic of M-stars are present in the 6800--7200\,\AA\ wavelength range.
    Bottom panel: The detection of an M-star donor is confirmed by the telluric-corrected, average VLT/FORS2 spectrum also obtained during quiescence. Both average spectra were dereddened using $E(B-V) = 0.6$ (see Sect.~\ref{sec:redd}), and the FORS2 spectrum was smoothed with a Savitzky-Golay filter, employing a window length of 9 data points and a third-order fitting polynomial for illustrative purposes.}
    \label{fig:quies_avg_spec}
\end{figure}

\subsection{\Ha\ emission radial velocity curves in quiescence from OSIRIS and FORS2}
\label{sec:Ha_RVC_above_quies}
Figure~\ref{fig:quies_avg_spec} (top) shows the average of the five OSIRIS spectra of \jj\ obtained on 1 Aug 2016, approximately one year after the data published by \cite{Lopez2019}. Notably, the OSIRIS spectrum exhibits features around 7000\,\AA\ that resemble those of mid-M dwarfs, although these are diluted due to significant slit contamination from the two much brighter nearby stars, caused by seeing conditions exceeding 1\arcsec. This is confirmed by the detection of an M-dwarf donor in the FORS2 quiescent spectrum.

The S/N of the individual OSIRIS spectra is sufficient for a radial velocity analysis based solely on the \Ha\ emission line from the accretion disc. Prior to measuring the velocities, we re-binned the spectra to a uniform velocity scale in a heliocentric frame and normalised them using a spline fit to the continuum using \texttt{MOLLY}\footnote{\url{https://cygnus.astro.warwick.ac.uk/phsaap/software/molly/html/INDEX.html}}. We determined the radial velocity of the \Ha\ emission line for each spectrum by cross-correlating its profile with a double-Gaussian template, following the method described by \cite{Schneider1980}. 

\begin{figure}
   \centering
   \includegraphics[width=0.96\columnwidth]{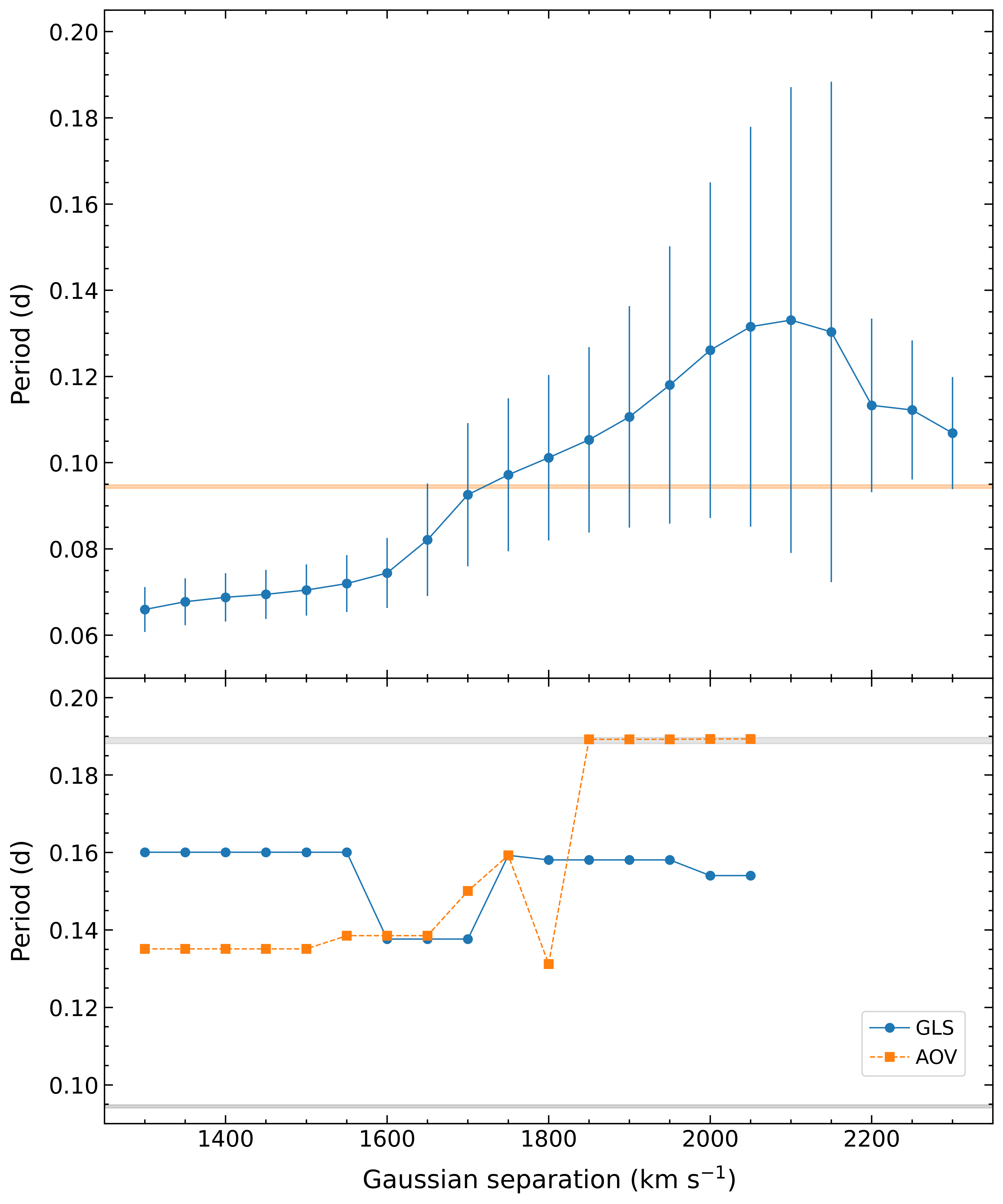}
   \caption{Top panel: Best sine-fit periods derived from the radial velocity curves of the \Ha\ emission line in the OSIRIS spectra, as a function of the double-Gaussian separation used in the velocity extraction. The shaded horizontal band indicates the periods derived from the light curves. Error bars represent 1$\sigma$ uncertainties. Bottom panel: Periods corresponding to the highest peaks in the GLS (blue circles) and AOV (orange squares) periodograms derived from the quiescent FORS2 \Ha\ radial velocity curves, as a function of the Gaussian separation used to extract the velocities. The period search was restricted to the range 0.068--0.25\;d. The lower shaded horizontal band marks the periods observed in the photometric light curves, while the upper band corresponds to twice these values.}
   \label{fig:j1910_mnras_fig_Ha_RVC}
   \end{figure}

We tested double-Gaussian separations ranging from 1300 to 2300\;\kms\ for a Gaussian FWHM of 300\;\kms, and fitted a sine function to each resulting radial velocity curve to analyse how the best-fit period varied with separation. This is illustrated in the top panel of Fig.~\ref{fig:j1910_mnras_fig_Ha_RVC}, which shows the best-fit period increasing from $0.066 \pm 0.05$ to $0.13 \pm 0.05$\;d. Beyond a separation of 2000\;\kms, the two Gaussians of the cross-correlation template begin to enter the continuum.

We repeated the analysis using Gaussian FWHMs ranging from 250 to 400\;\kms\ in steps of 50\;\kms\ and found no significant differences.
Given the limited number of radial velocity measurements (five data points) spanning only $\approx 0.095$\;d, period determination is severely constrained. Although sine fitting can in principle be applied, the small number of observations provides little redundancy and renders the solution highly sensitive to noise and outliers. Furthermore, the short time baseline leads to strong degeneracies between period and phase, increasing the risk of aliasing and spurious fits. As such, reliable period recovery from the OSIRIS spectra alone is not feasible.

We followed the same procedure for the quiescent FORS2 spectra. We subsequently computed GLS and AOV periodograms from the resulting radial velocity curves scanning periods between 0.068 and 0.25\;d. The highest-peak periods vary with Gaussian separation and have approximate values of 0.14, 0.16, and 0.19\;d (Fig.~\ref{fig:j1910_mnras_fig_Ha_RVC}, bottom panel). The latter happens to be about twice the typical periodicity observed in the light curves; however, we deem the time sampling of our \Ha\ radial velocity curve during quiescence too scarce to conduct a reliable period search. In fact, from the minimum time separation between two consecutive samples in our data, $\Delta t$, the pseudo-Nyquist period is $2 \times \mathrm{min}(\Delta t) \approx 0.068$\;d. For non-uniform sampling, the effects of aliasing and windowing prevent reliable recovery of periods close to this value \citep{VanderPlas2018}, and candidates shorter than about $(4-5) \times \mathrm{min}(\Delta t) \approx 0.14 - 0.17$\;d should not be trusted. This places the observed highest-peak periods at the edge of reliable detectability.

\begin{table}
 	\centering
 	\caption{Comparison of the average FWHM$_0$ and EW of the \Ha\ line measured in the spectra of \jj~taken during outburst and quiescence. 
    }
 	\label{tab:comparison}
        \begin{threeparttable}
 	\begin{tabular}{ccc} 
     \toprule\noalign{\smallskip}
                      & FWHM$_0$ (\kms)  & EW (\AA) \\
    \midrule\noalign{\smallskip}
    Outburst        & \multirow{2}{*}{$555\pm30$} & \multirow{2}{*}{$2.07\pm0.03$}  \\
    (WHT/ACAM)      &                            & \\\midrule\noalign{\smallskip}
    Quiescence & \multirow{2}{*}{$(960\pm90)^*$} & \multirow{2}{*}{$(17\pm2)^*$} \\
    (GTC/OSIRIS)     &            &          \\\midrule\noalign{\smallskip}
    Quiescence    & \multirow{2}{*}{$1001 \pm 11$} & \multirow{2}{*}{$91 \pm 1$}  \\
    (VLT/FORS2)    &  &   \\
    \bottomrule\noalign{\smallskip}
 	\end{tabular}
\begin{tablenotes}\footnotesize
\item[*] The values derived from the OSIRIS spectra are unreliable due to flux contamination from two brighter stars located just south of \jj, caused by seeing conditions.
\end{tablenotes}
\end{threeparttable}
\end{table}

\subsection{Constraints on reddening from DIBs}
\label{sec:redd}
 \begin{table}
 	\centering
 	\caption{EW and \ebv\ obtained for selected interstellar features.}
 	\label{tab:ew}
 	\begin{tabular}{ccc} 
     \toprule\noalign{\smallskip}
 		Interstellar & EW & \ebv \\
 	    feature	 & (\AA) &       \\
     \midrule\noalign{\smallskip}
 		5780\,\AA-DIB & $0.34\pm0.03$ & $0.64\pm0.06$ \\
      6614\,\AA-DIB & $0.14\pm0.01$ & $0.61\pm0.05$ \\
 		8621\,\AA-DIB & $0.20\pm0.03$ & $0.54\pm0.08$ \\
 		\Ion{Na}{i} doublet & $2.6\pm0.2$ & $>0.5$ \\
     \bottomrule\noalign{\smallskip}
 	\end{tabular}
 \end{table}
 
In the absence of a \textit{Gaia} DR3 \citep{GaiaDR3} counterpart and thus no parallax for \jj, we estimate the reddening using diffuse DIBs. We excluded those affected by strong blending or poorly defined continua---we focused on 5780, 6614, and 8621\;\AA. The reddening was derived from the measured EWs of these DIBs using the empirical relations from \cite{Munari2008} and \citet{Lan2015}, and \cite{Krelowski2019} for the 5780-\AA\ band. The results are presented in Table~\ref{tab:ew}.

The values obtained from the interstellar \Ion{Na}{i} doublet and the 8621-\AA\ band provide lower bounds and are consistent with the reddening inferred from the 6614-\AA\ DIB, as is the value from the 5780-\AA\ band. Additionally, the DIBs at 5780 and 6614\,\AA\ can be used to estimate the hydrogen column density ($N_\mathrm{H}$) following the empirical relations of \citet{Lan2015}, serving as a consistency check for our reddening measurements. We find $N_\mathrm{H} = (3.9 \pm 1.3) \times 10^{21} \,\mathrm{cm}^{-2}$, in agreement with values of $N_\mathrm{H} \approx (3.5 - 4.0) \times 10^{21} \,\mathrm{cm}^{-2}$ derived from X-ray spectroscopy during outburst by \citet{Degenaar2014} and \citet{Nakahira2014}. Using the relation from \cite{Guver2009}, $N_\mathrm{H} = 2.21 \times 10^{21} \, A_V$, and adopting $R_V = 3.1$, these values correspond to $E(B-V) \approx 0.51$--0.58, in line with our results. We finally adopt \ebvval{0.6} for \jj.

The adopted reddening value corresponds to $E(g - r)$ values of 0.68 and 0.61 when using the Bayestar19-recommended conversion, $E(B - V) = 0.884 \, E(g - r)$, or $E(B - V) = 0.981 \, E(g - r)$  \citep{Schlafly2011}, respectively. However, the Bayestar19 dust map along the line of sight to \jj\ saturates at $E(g - r) \approx 0.58$, which may reflect limitations in the 3D dust map's ability to trace dense or complex sight lines.
This is particularly true in regions of high extinction or low Galactic latitude, where dust maps are known to saturate or under-predict reddening due to limited background stars \citep[see e.g.][]{Green2019}.

\begin{figure}
	\includegraphics[width=0.97\columnwidth]{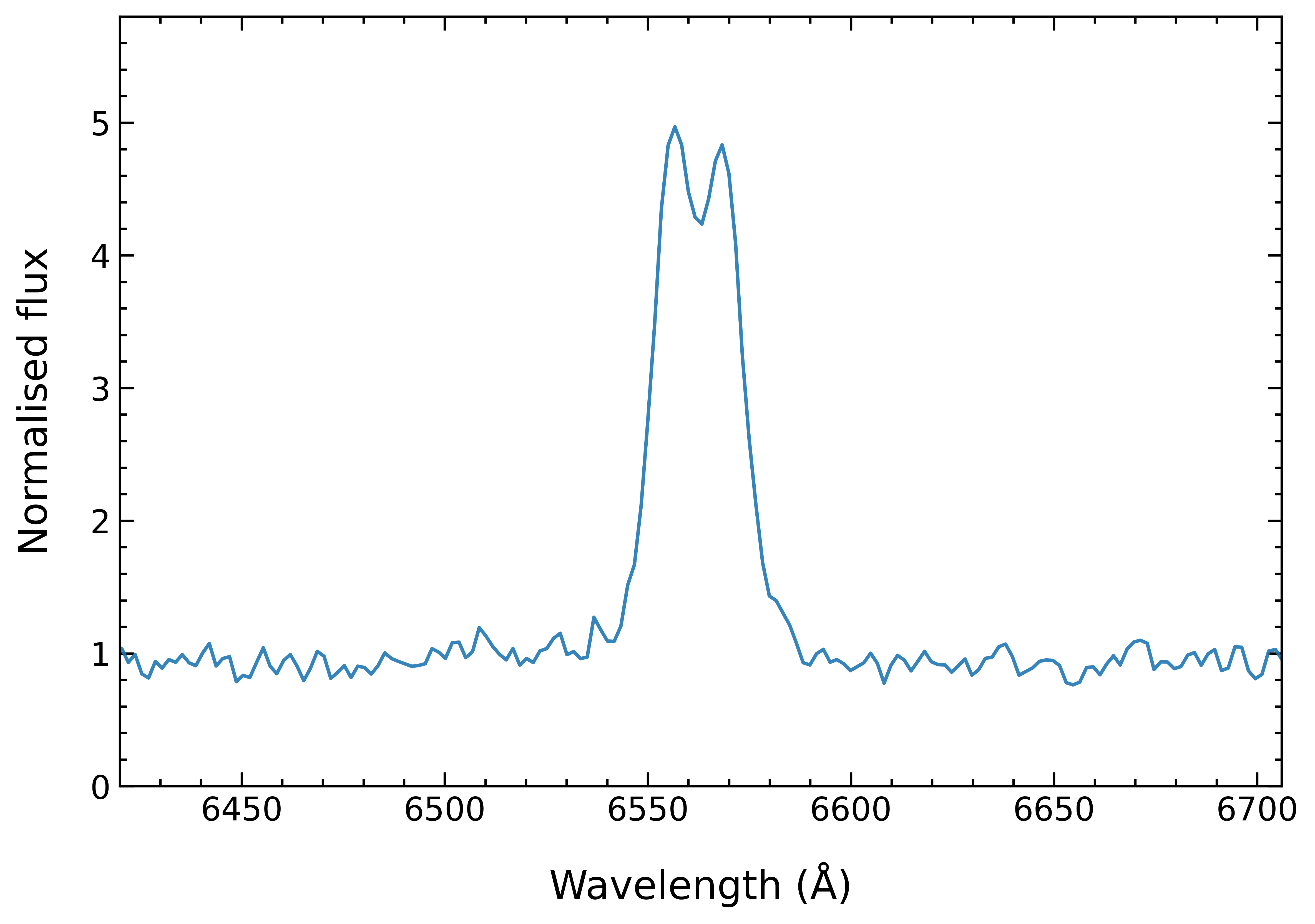}
    \caption{Average of the 15 FORS2 spectra obtained during quiescence. The profile of the H$\alpha$ emission line is double-peaked.}
    \label{fig:fors2_Ha_DPeak}
\end{figure}

 \begin{figure*}
\sidecaption
  \includegraphics[width=12cm]{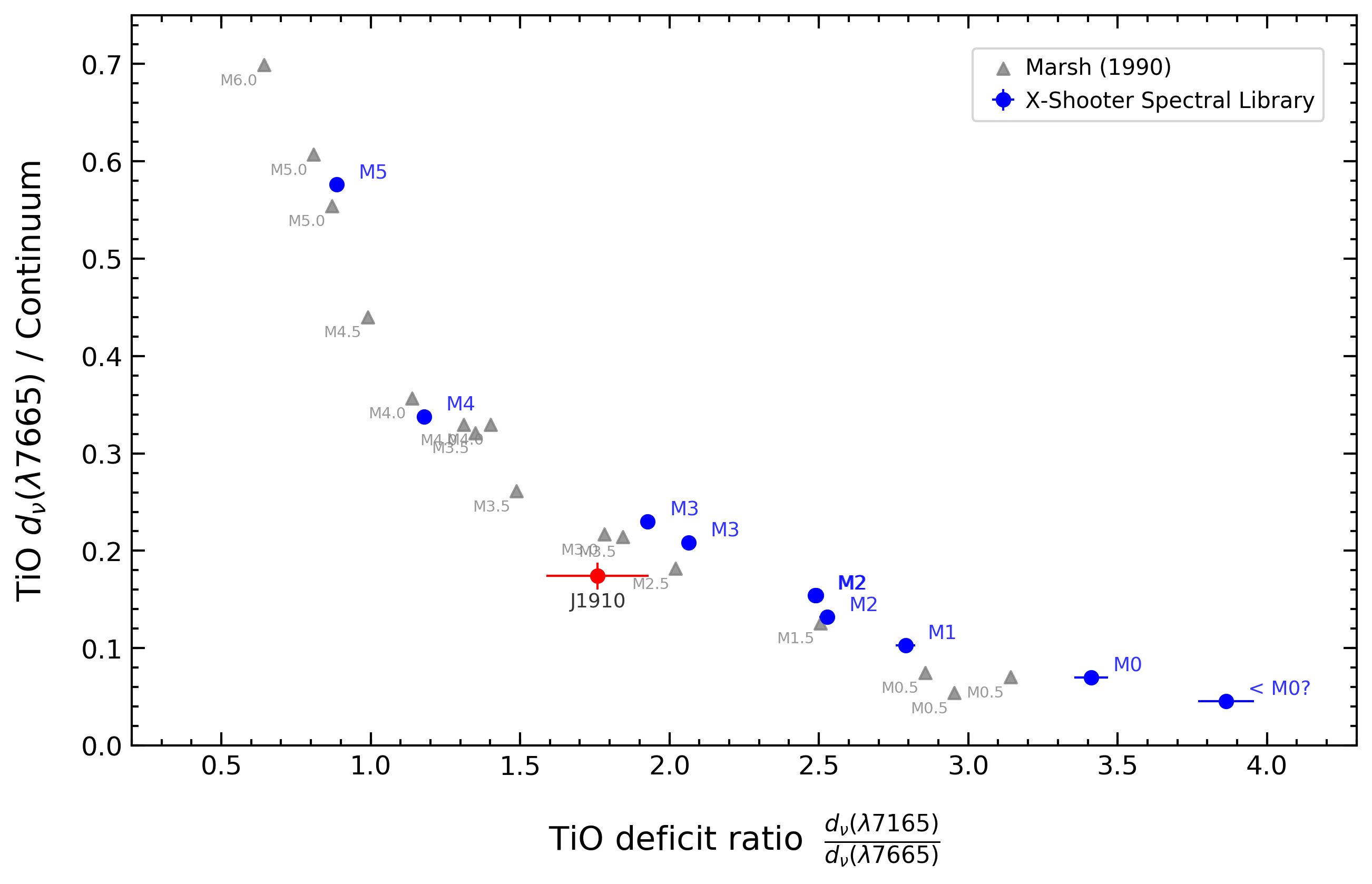}
     \caption{Fractional depth of the TiO \textlambda7665 band as a function of the TiO band deficit ratio for \jj\ and comparison M-type stars. Blue dots indicate X-Shooter Library M dwarfs, while grey triangles correspond to those used by \cite{Marsh90}.\\}
     \label{fig:TiO_flux_deficits}
\end{figure*}

\subsection{Detection of the donor star in quiescence: spectral type and distance}
\label{sec:spec_typing}

We present the dereddened average spectrum of \jj\ computed from the VLT/FORS2 data obtained during quiescence in the bottom panel of Fig.~\ref{fig:quies_avg_spec}. For the averaging process, we included only the spectra where the donor star is more prominently detected (spectra 8, 12, 13, and 14 in Table \ref{tab:FORS2_quies}). 

In the wavelength range sampled by our FORS2 data, the spectrum of \jj\ in quiescence is dominated by the flux from an M-type star and broad \Ha\ emission ($\mathrm{FWHM}_0 \approx 1000$\;\kms; Table~\ref{tab:comparison}). The observed $\mathrm{FWHM}_0$ of \Ha\ during quiescence is too large to originate from the companion star, whose \Ha\ emission typically exhibits FWHM values of only a few hundred \kms\ (e.g. \citealt{rattietal12-1} for a neutron star X-ray binary and \citealt{rodriguez-giletal15-1,rodriguez-giletal20-1} for nova-like cataclysmic variables in the low state). The large width and double-peaked morphology of the line are instead consistent with an accretion disc origin (see Fig.~\ref{fig:fors2_Ha_DPeak}).

The TiO band head at 7165\,\AA\ is distinctly evident, while that at 7665\,\AA\ is also prominently observed. \cite{Young+Schneider81} and \cite{Wade+Horne88} demonstrated that the TiO band ratio, 7165\,\AA /7665\,\AA, is sensitive to spectral type.  
Following the latter authors, we fitted a first-order polynomial to the continuum over the wavelength ranges \textlambda\textlambda 7450--7550 and \textlambda\textlambda 8130--8170 and calculated the flux deficits using the equations:

\begin{equation*}
d_\nu(\lambda7165) = \frac{\int_{7140}^{7190} (c_\lambda - f_\lambda) \, d\lambda / \lambda}{\int_{7140}^{7190} d\lambda / \lambda}
\end{equation*}

\noindent
and

\begin{equation*}
d_\nu(\lambda7665) = \frac{\int_{7640}^{7690} (c_\lambda - f_\lambda) \, d\lambda / \lambda}{\int_{7640}^{7690} d\lambda / \lambda}~,
\end{equation*}

\noindent
where $f_\lambda$ is the flux density of the spectrum and $c_\lambda$ represents the flux density of the linear fit to the selected continuum regions. We computed both the ratio of these deficits to the continuum flux at 7500 \AA\ and the deficit ratio for \jj\ and selected flux-calibrated and dereddened M-dwarf spectra from the X-Shooter Spectral Library (XSL) DR3 \citep{Verro2022}, spanning the spectral types M0 to M5\footnote[6]{HD~111631 (M0), HD~209290 (M0), HIP~96710 (M1), HD~119850 (M2), HIP~75423 (M2), GJ~109 (M3), HIP~84123 (M3), HD~125455B (M4), and GJ~866 (M5)}, after degrading to match the FORS2 resolution.
For \jj, we obtain a TiO deficit ratio of $1.76 \pm 0.17$ and a fractional depth of the TiO \textlambda7665 band of $0.17 \pm 0.01$, with the 1$\sigma$ uncertainties derived from 10\,000 Monte Carlo simulations. The results are illustrated in Fig.~\ref{fig:TiO_flux_deficits}, which also includes results from \cite{Marsh90}. The spectral types of the observed M-type stars reported in that study were adopted from \cite{Boeshaar76}, with additional comparison points included from \cite{Wade+Horne88}. We revised the Boeshaar spectral types adopted by \citeauthor{Wade+Horne88} using updated and consistent classifications retrieved from the literature via SIMBAD \citep{Wenger2000}: Gl~328 (M0), Gl~381 (M2.5), Gl~273 (M3.5), Gl~299 (M4.5), and Gl~285 (M4.5).
We were unable to revise the spectral types of the M-type stars observed by \cite{Marsh90} because their names were not reported and the observing logs are unavailable.

To estimate the spectral type of the \jj\ donor, we performed linear fits of the TiO deficit ratio against spectral type in the deficit ratio interval 1.05--2.60. A spectral type of approximately M3–3.5 is obtained when fitting the XSL data, the revised data from \cite{Marsh90}, or a combination of both.

Depending on its fractional contribution to the total flux, the continuum light from the accretion disc in \jj\ shifts it below the line defined by M-type stars, while the TiO band deficit ratio remains consistent unless the disc itself produces TiO bands. The fractional depth of the TiO \textlambda7665 band in \jj\ is below the values observed in single dwarfs with similar TiO band deficit ratios but remains close to the M-dwarf sequence, indicating a significant contribution of the donor star to the total flux of the system. We conducted linear fits of the deficit ratio against fractional depth for the same three data sets as above. By comparing the predicted fractional depth with the measured value, the donor star in \jj\ is found to contribute approximately 70\% to the total flux in the wavelength range considered.

To derive the distance, we used the observed average $I$-band magnitude from the FORS2 acquisition images (Table~\ref{tab:FORS2_quies}), $m_I = 21.52 \pm 0.01$, and assumed that 70\% of the flux in this band originates from the donor. The colour excess of $E(B - V) = 0.6$ translates to an extinction of $A_V = 3.1 \, E(B - V) = 1.86$, and the extinction in the $I$-band was then computed as $A_I \approx 0.479 \, A_V = 0.89$ \citep{Cardelli1989}. Applying the distance modulus and correcting for extinction and relative flux contribution of the donor, we obtain a distance to \jj\ between 2.8 and 4.0\;kpc. This would place the system at a height above the Galactic plane of $\left|z\right| \simeq 0.4$\,kpc.

\subsection{Constraints on the binary parameters from the \Ha\ profile}
\label{sec:binparam}
Empirical relationships developed from studies of quiescent X-ray transients allow us to estimate key binary system parameters, such as the velocity semi-amplitude of the donor star ($K_2$), the binary mass ratio ($q$), and the orbital inclination ($i$), by analysing the double-peaked \Ha\ emission line (Fig.~\ref{fig:fors2_Ha_DPeak}) from the accretion disc \citep[][]{Casares2015, Casares2016, Casares2022}. 

We applied these relationships using the normalised FORS2 spectra. 
To mitigate the possible influence of the TiO band depression on the red side of \Ha, we rebinned the spectra to the 6300--6800\;\AA\ range and re-normalised them using a third-order spline fit to the continuum.
We then fitted both single-Gaussian and symmetric two-Gaussian models to the individual spectra and the average spectrum (Fig.~\ref{fig:fors2_Ha_DPeak}) over a velocity range of $\pm 10\,000$\;\kms\ centred on \Ha, adjusting the models to match the instrumental resolution. The single-Gaussian fit provided the FWHM, while the two-Gaussian fit yielded the FWHM of each Gaussian component ($W$) and the double-peak separation ($DP$)---see Fig.~4 of \citet{YanesRizo2025} for a graphical description of these emission‑line properties.

The measurement of $K_2$ depends solely on the $\mathrm{FWHM}_0$ of the \Ha\ profile, and can be  determined from individual spectra through the following relation \citep{Casares2015}:
\begin{equation}
    K_2 = 0.233\,(13) \times \text{FWHM}_0~.
    \label{equ:k2}
\end{equation}

We used $\mathrm{FWHM}_0 = 990 \pm 45$\;\kms, the median and the standard deviation of the distribution of individual measurements. Here, the uncertainty in  FWHM$_0$ properly reflects orbital variations. Using this FWHM$_0$ estimate and its uncertainty, and performing a Monte Carlo simulation with $10^5$ realisations of Eq.~\ref{equ:k2}, we obtain $K_2 = 230 \pm 17$\;\kms.

The binary mass ratio ($q$) can be estimated using the correlation from \citet{Casares2016}, which relates $q$ to the $DP$-to-FWHM$_0$ ratio:
\begin{equation}
    \log q = -6.88\,(0.52) - 23.2\,(2.0) \log \left( \frac{DP}{\mathrm{FWHM_0}} \right)~.
    \label{equ:q}
\end{equation}

\noindent
To estimate $q$, we applied a symmetric two-Gaussian fit to the weighted average spectrum, yielding $q = 0.032^{+0.013}_{-0.009}$. As shown in \cite{Casares2016}, this approach is necessary because individual spectra typically have insufficient S/N for reliable analysis. Moreover, averaging across the orbit helps to minimise distortions from features such as hot spots or disc eccentricities, which could otherwise skew the estimation of $q$.

The single-Gaussian fit to the average line profile gives $\mathrm{FWHM_0} = 1001 \pm 11$\;\kms, while the symmetric two-Gaussian fit yields $DP = 586 \pm 5$\;\kms\ and $W = 506 \pm 9$\;\kms. The value of $W$ will be used to estimate the orbital inclination of the system in the following analysis.  

The orbital inclination can be estimated from the depth of the trough ($T$) between the two peaks of the \Ha\ emission profile using the relation from \cite{Casares2022}:
\begin{equation}
    i\;[^\circ] = 93.5\,(6.5)\,T + 23.7\,(2.5)~,
    \label{equ:i}
\end{equation}
where $T$ is defined as $1 - 2 \sqrt{1 - \left( DP/W \right)^2}$. We corrected $T$ for the effect of instrumental resolution as described in appendix~A of \cite{Casares2022}.

\noindent
Using the values of $DP$ and $W$ calculated earlier, we obtain $T = 0.27 \pm 0.03$ and an inclination of $i={49^\circ} \pm 4^\circ$, with uncertainties derived from a Monte Carlo simulation. 
We verified that the double-peaked \Ha\ profile is affected by contamination from the nearby star also on the slit, with the contamination level depending on seeing. Repeating the analysis using only the average of the five spectra showing less than 1\% contamination within 1-$\sigma$ of the spatial profile of \jj, as determined from two-Gaussian fits to the collapsed spectra near \Ha, favours a lower inclination ($i \approx 40^{\circ}$); however, this result should be interpreted with caution owing to the incomplete orbital phase coverage of these spectra. Nevertheless, in Sect.~\ref{sec:discuss_i} we show that the orbital inclination is likely much lower.

\section{Discussion}
\label{sec:discuss}
\subsection{Optical light curve morphology and the orbital period}\label{subsec:P}

Our GLS search of the combined 2012 SMARTS, WFC, and Fresno SRO light curves identifies a strong signal at $2.26(2)\;$h. At first sight, the photometric variability could be attributed to X-ray heating of the donor star, but several arguments disfavour this scenario in \jj. In particular, a very low mass ratio ($q\!\lesssim\!0.07$) means that the flared outer disc shadows the donor from direct X-rays, so irradiation is inefficient, as argued by \cite{Torres2021} for the very similar MAXI~J1659--152. 
Nevertheless, if X-ray heating of the donor were significant, the low orbital inclination would make its photometric detection unlikely \citep{Haswell2001}. Additionally, no narrow emission-line components attributable to an illuminated donor are seen in our spectra. Together, these points indicate an accretion-disc origin for the photometric modulation.

\begin{figure*}
	\includegraphics[width=1.02\columnwidth]{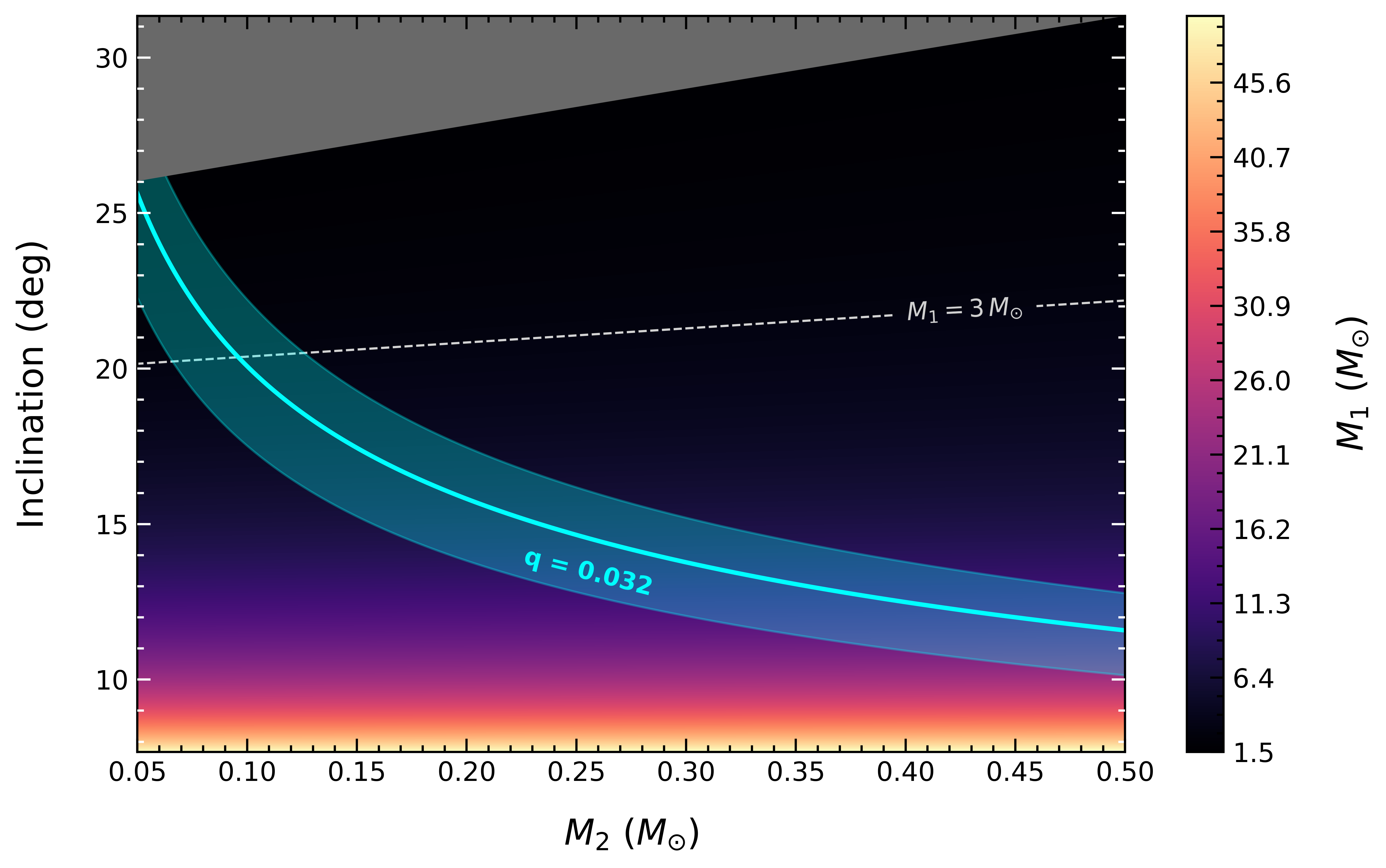}\includegraphics[width=1.02\columnwidth]{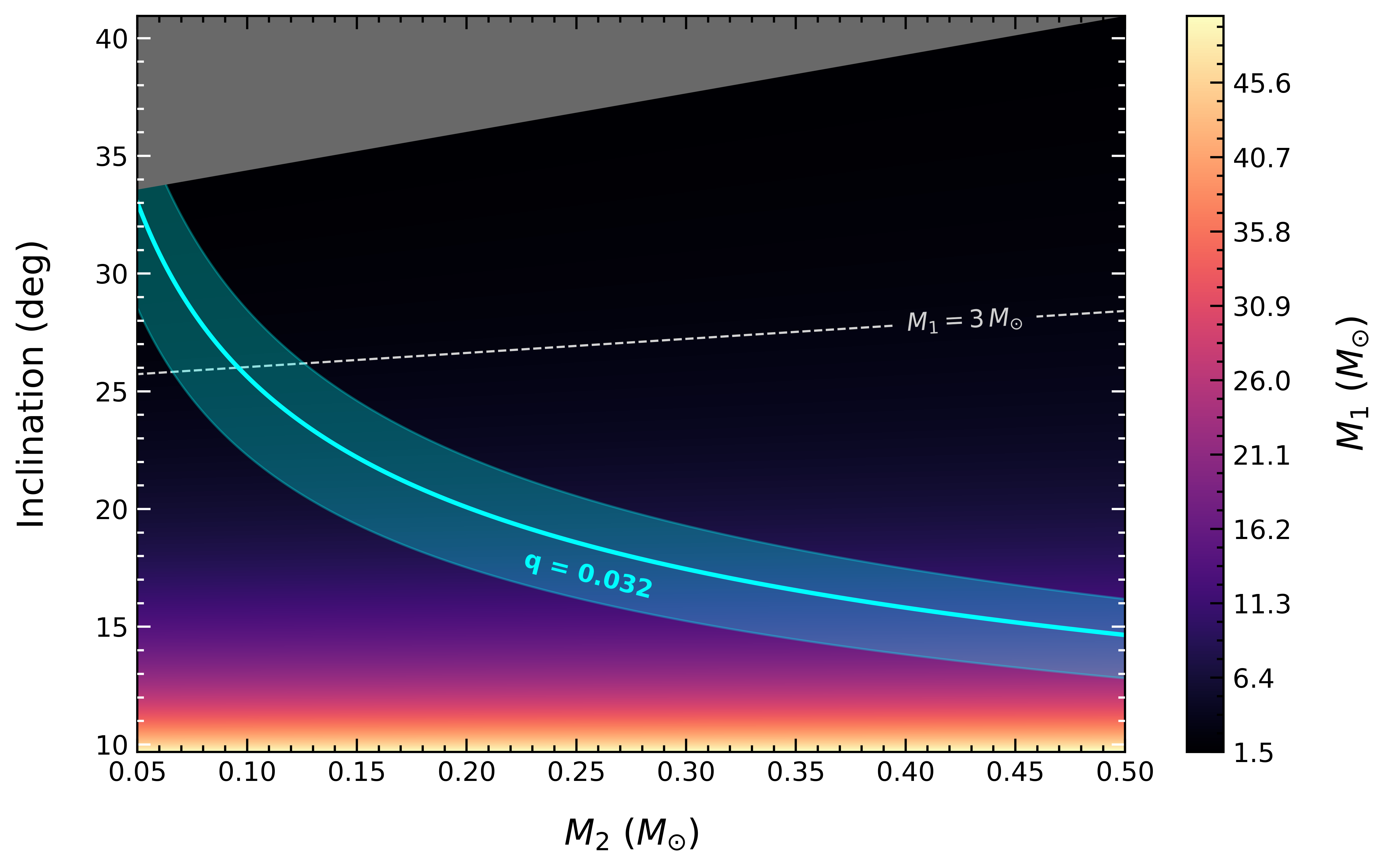}
    \caption{Relationship between the orbital inclination and the mass of the donor star ($M_2$), with the mass of the compact object ($M_1$) represented by colour, for orbital periods of 2.25 (left) and 4.50\;h (right). The solid cyan line and cyan shaded area correspond to a mass ratio of 0.032 and its uncertainty, respectively. The light grey dashed line marks  $M_1 = 3$\;\Msun, and the dark grey triangular areas correspond to $M_1 < 1.5$\;\Msun.}
    \label{fig:M2_i_M1}
\end{figure*}

The profile of the \jj\ light curves (Figs.~\ref{j1910_mnras_fig_GLS_01} and \ref{fig:j1910_mnras_fig_GLS_june_july_2012}) suggests that the waveform might be a double-wave (i.e. two-humped), with a fundamental period of $\simeq\!4.52$\;h. Independent support for this possibility comes from the FORS2 \Ha\ radial-velocity wings. For Gaussian separations $\gtrsim\!1850$\;\kms, the AOV periodograms show excess power near 4.54\;h ($0.189$\;d; Fig.~\ref{fig:j1910_mnras_fig_Ha_RVC} bottom panel). If this interpretation is correct, the double-wave modulation observed in \jj\ may well be an ``early superhump'', a common feature of WZ\;Sge-type dwarf novae during outburst.

In WZ\;Sge dwarf novae, it is thought that double-humped modulations arise when the outer disc reaches the 2:1 resonance, producing a two-armed, vertically extended pattern that modulates the optical flux twice per orbit \citep[e.g.][]{Kato2002, Uemura2012}. This configuration is only attainable for extreme mass ratios, $q \lesssim 0.08$, because the 2:1 resonance radius must lie within the tidal truncation radius. Our spectroscopic estimate for \jj, $q = 0.032 \pm 0.010$, comfortably satisfies this criterion, so the observed double-wave profile could be explained within the early-superhump framework.

The $R$-band time series of MAXI~J1659--152 presented by \cite{Kuroda2010}\footnote{\href{http://maxi.riken.jp/FirstYear/ppt/P04KURODA.pdf}{\url{http://maxi.riken.jp/FirstYear/ppt/P04KURODA.pdf}}} also exhibits a double-humped profile with an amplitude of $\approx\!0.04$\;mag
that persists for at least the first 18\;d after the optical detection of the transient. This modulation is identified by \cite{Torres2021} as a possible early superhump. Analogous behaviour was observed in the black hole transient XTE\;J1118+480: its 2000--2001 outburst displayed a 4.09-h double-wave superhump just 0.3\% longer than its 4.065-h orbital period \citep{Uemura2000, Zurita2002}, while the cataclysmic variable V503\;Cyg alternates between single- and double-humped profiles within a single superoutburst \citep{Harvey1995}.

Both the amplitude and the lifetime of these modulations align with the canonical properties of early superhumps in WZ\;Sge systems during outburst: most systems display modulation amplitudes $\lesssim\!0.05$\;mag, with occasional extremes up to 0.35\;mag \citep{Kato2015}, and the signal typically fades within $\sim\! 15$\;d of outburst onset \citep[their table 9]{Nakata2013}. The $r$-band amplitude of the modulation in \jj\ is about 0.02\;mag, in line with early superhumps, but persisted for at least 46\;d.
Additionally, the relatively low amplitude of the modulation also supports a low inclination. As shown by \cite{Haswell2001}, the energy released by changes in the disc due to superhumps is negligible compared to the X-ray irradiation during outburst. Thus, detectable superhumps typically require additional asymmetries, such as tilted or warped discs, as invoked by \cite{Thomas2022} for MAXI\;J1810$+$070. The modest modulation observed in \jj\ suggests no such enhancement, reinforcing the interpretation of a low-inclination geometry (see Sect.~\ref{sec:discuss_i}).

However, a reliable determination of the orbital period from the radial velocity curves obtained with ACAM, OSIRIS, or FORS2 was not possible. Consequently, it remains unclear whether the orbital period is close to 2.26\;h or approximately twice that value. Using the donor spectral type-orbital period relations from \cite{Smith+Dhillon1998}, we estimate spectral types of $\mathrm{M5}\!\pm\!1$ and $\mathrm{M2}\!\pm\!4$ for the shorter and longer periods, respectively. These estimates are too uncertain to meaningfully constrain the orbital period, especially when compared to our measured spectral type of M3--M3.5.

We computed superhump excesses, $\epsilon$, using the $q$-$\epsilon$ relationships from \cite{McAllister2019}. For our mass ratio $q=0.032 \pm 0.010$, we obtain orbital periods of approximately 2.25 and 4.50\;h for superhump periods of 2.26 and 4.52\;h, respectively, which are slightly shorter than the corresponding superhump period in both cases.

We can estimate the masses of the compact object and the donor star in \jj~using:
\begin{equation}
\label{equ:fm}
    \frac{M_1 \sin^3 i}{(1 + q)^2} = \frac{P K_2^3}{\rm 2\pi G}~,
\end{equation}

\noindent
where $M_1$ is the mass of the compact object and $q=M_2/M_1$, with $M_2$ the mass of the donor star.
With our derived values of $K_2$, $q$, $i$, and $P = 2.26(2)$\;h, a Monte Carlo simulation yields $M_1\! =\! 0.30^{+0.10}_{-0.07}$\;\Msun\ and $M_2\! =\! 0.009^{+0.005}_{-0.004}$\;\Msun. For twice that period, the inferred values are doubled. The derived masses for both the compact object and the donor are clearly too low. This suggests that either the true orbital period is $\sim\!30\!-\!40$ times longer ($\sim\!3\!-\!4$\;d), or---more likely---the orbital inclination has been overestimated.

\subsection{Inclination constraints and compact object mass}\label{sec:discuss_i}

From semi-empirical donor sequences for Roche-lobe-filling stars in cataclysmic variables \citep[e.g.][]{Knigge2011}, the donor's observed spectral type of M3--3.5 suggests a (conservative) mass of $M_2\! =\! 0.25$--$0.35$\,\Msun. To constrain $M_1$ and $i$, we performed a Monte Carlo simulation incorporating uncertainties in the orbital period, radial velocity semi-amplitude, and mass ratio.

We adopted $K_2\! =\! 230\! \pm\! 17$\;\kms, a mass ratio $q = 0.032 \pm 0.010$, and considered two orbital period scenarios: $P = 2.25 \pm 0.01$\;h, and twice that value ($P = 4.50 \pm 0.03$\;h).

For each of $10^5$ trials, we computed the mass function (Eq.~\ref{equ:fm}) and solved for $M_1$ and $i$.

The resulting posterior distributions yield the following median values and 68\% confidence intervals:
\begin{itemize}
  \item Shorter period ($P = 2.25$\;h):
  \begin{itemize}
    \item[]
    \item[] $M_2 = 0.25$\;\Msun:~~ $M_1 =~~ 8^{+3}_{-2}$\;\Msun,~ $i = 14{^\circ} \pm 2^\circ$
    \item[] $M_2 = 0.35$\;\Msun:~~ $M_1 = 11^{+5}_{-3}$\;\Msun,~ $i = 13{^\circ} \pm 2^\circ$\\
  \end{itemize}

  \item Longer period ($P = 4.50$\;h):
  \begin{itemize}
    \item[]
    \item[] $M_2 = 0.25$\;\Msun:~~ $M_1 =~~ 8^{+3}_{-2}$\;\Msun,~ $i = 18{^\circ} \pm 3^\circ$
    \item[] $M_2 = 0.35$\;\Msun:~~ $M_1 = 11^{+5}_{-3}$\;\Msun,~ $i = 16{^\circ} \pm 
    2^\circ$
  \end{itemize}
\end{itemize}

In both scenarios, the compact object mass falls within the black hole regime, and the derived inclinations are low. 
In Fig.~\ref{fig:M2_i_M1}, we present the mass of the compact object as colours for a range of inclinations and donor masses, for the two orbital periods considered.

Evidence for a low inclination comes from the following: \cite{Degenaar2014} reported the absence of ionised disc winds during the softest state of the X-ray outburst, along with a relatively low inner disc temperature ($kT \simeq 0.5$\;keV), in contrast to the $kT\! >\! 1$\;keV temperatures typically found in black hole systems; \cite{MunozDarias2013} showed that such cooler disc temperatures are systematically observed at lower inclinations, due to relativistic effects. Additionally, the lack of disc winds is consistent with a low inclination \citep{Ponti2012}. Together, these findings favour a low orbital inclination for \jj, consistent with our results based on the measured mass ratio and donor mass range inferred from its observed spectral type.

\subsection{Alternative scenario: A neutron star accretor}
Assuming a canonical neutron star mass of $M_1\!\approx\!1.4$\;\Msun\ and that the M3-type donor is a main-sequence star with a mass of $M_2\!\approx\!0.3$\;\Msun, the resulting mass ratio would be $q\!\approx\!0.21$. This value remains consistent with the observed photometric modulation being due to a superhump---though not an early superhump. In this scenario, the orbital period would be $\approx\!2.25$\;h, and using the FWHM$_0$-derived $K_2$, the system inclination would be approximately $30{^\circ}$. A challenge to this interpretation comes from the $q$ and $i$ values inferred through empirical correlations with the double-peaked emission lines. One further point worth mentioning is that, under the alternative scenario of a 4.5-h orbital period, the Roche-lobe density constraint suggests the M3-type donor could be a normal main-sequence star. In contrast, for a 2.25-h period, the donor would need to be significantly cooler, implying it likely descended from a more massive star that began Roche-lobe overflow while already significantly evolved---similar to the evolutionary histories proposed for XTE\;J1118+480 \citep{Haswell2002} or the cataclysmic variable SDSS\;J170213.26+322954.1 \citep{Littlefair2006}.

\section{Conclusions}
\label{sec:con}
In this paper, we have presented a comprehensive study of \sw\ based on time-series optical photometry and spectroscopy obtained during its 2012 outburst, subsequent decline, and quiescent state. 
We have examined the system's orbital and stellar parameters, with particular emphasis on constraining the orbital period, inclination and component masses.

Our analysis of the FORS2 quiescent optical spectrum, dominated by molecular absorption bands characteristic of M dwarfs, enables a direct spectral classification of the donor star as M3--3.5. It further suggests that the donor contributes approximately 70\% of the red optical flux. Based on the inferred spectral type and the measured reddening \ebv~$=0.60\pm0.05$, derived from the analysis of DIBs, we estimate a system distance of between $2.8$ and $4.0$\;kpc.

Time-resolved photometry during outburst reveals a  photometric modulation with a period of $2.26(2)$\;h, consistent with an early superhump as observed in extreme mass-ratio accreting systems. This introduces an ambiguity regarding the true orbital period, which may be slightly shorter than twice this photometric period.

The observed double-peaked \Ha\ emission line profile provides an estimate of the binary mass ratio, $q\!=\!0.032^{+0.013}_{-0.009}$, and a radial velocity semi-amplitude of the donor of $K_2=230^{+17}_{-16}$\;\kms. However, the inclination derived from the H$\alpha$ profile, $i = 49^\circ \pm 4^\circ$, 
may be too high compared to independent X-ray-based estimates. 

To further constrain the inclination and the mass of the compact object, we considered two possible orbital period scenarios, $2.25$ and $4.50$\;h, and adopted a conservative donor mass range consistent with the measured spectral type ($M_2\!=\!0.25-0.35$\;\Msun). Our Monte Carlo simulations yield a low inclination in the range $i\!=\!13^\circ\!-\!18^\circ$ and $M_1\!=\!8\!-\!11$\;\Msun. We favour the 4.5-h orbital period and a black hole accretor in \jj, but a neutron star cannot be discarded. 

Future time-resolved spectroscopy under excellent seeing conditions ($\lesssim\!0.5$\arcsec) complemented by $r$- and/or $i$-band light curves will be crucial to 
conclusively determine the orbital period and the true nature of its compact object.

\begin{acknowledgements}
We thank the referee for a prompt and positive report, and for the constructive suggestions that helped improve the manuscript.\\
We are grateful to Tom Marsh for the use of \verb|molly|. Beyond his scientific brilliance, Tom was a generous colleague and dear friend. His absence is profoundly felt.\\
PR-G, MAPT, and TM-D acknowledge support by the Agencia Estatal de Investigación del Ministerio de Ciencia e Innovación (MCIN/AEI) and the European Regional Development Fund (ERDF) under grant PID2021--124879NB--I00. DMS acknowledges support via a Ramón y Cajal Fellowship RYC2023--044941. PGJ~is supported  by the European Union (ERC, Starstruck, 101095973). Views and opinions expressed are however those of the author(s) only and do not necessarily reflect those of the European Union or the European Research Council Executive Agency. Neither the European Union nor the granting authority can be held responsible for them. PAC acknowledges the Leverhulme Trust for an Emeritus Fellowship. PS and DMR are supported by Tamkeen under the NYU Abu Dhabi Research Institute grant CASS. This research has made use of the SIMBAD database, the VizieR catalogue access tool, and Aladin sky atlas (CDS, Strasbourg Astronomical Observatory, France).
This research has also made use of \verb|IRAF|, which is distributed by the National Optical Astronomy Observatory, which is operated by the Association of Universities for Research in Astronomy (AURA) under a cooperative agreement with the National Science Foundation. This research is based on data obtained from the Astro Data Archive at NSF’s NOIRLab. These data are associated with observing program NOAO 2012A--0424 (PI C. Britt). NOIRLab is managed by the Association of Universities for Research in Astronomy (AURA) under a cooperative agreement with the National Science Foundation.
Based on observations collected at the European Organisation for Astronomical Research in the Southern Hemisphere under ESO programme(s) 097.D--0270 (PI P. Jonker), 
and observed with GTC under program GTC38-16B (PI J. Casares).
The WHT and INT are operated by the Isaac Newton Group of Telescopes. These telescopes together with the GTC are installed at the Spanish Observatorio del Roque de los Muchachos of the Instituto de Astrof\'isica de Canarias, on the island of La Palma.
This research used photometry taken at Fresno State's station at Sierra Remote Observatories. We thank Dr. Greg Morgan, Dr. Melvin Helm, and Dr. Keith Quattrocchi, and the other SRO observers for creating this fine facility.

This research made use of 
APLpy \citep{Robitaille2012}, 
SciPy \citep{Virtanen2020} and \href{http://www.astropy.org}{AstroPy} \citep{Astropy2022}.
\end{acknowledgements}

\bibliographystyle{bibtex/aa}
\bibliography{bibtex/biblio_j1910}

\begin{thebibliography}{88}
\expandafter\ifx\csname natexlab\endcsname\relax\def\natexlab#1{#1}\fi

\bibitem[{{Astropy Collaboration} {et~al.}(2022){Astropy Collaboration}, {Price-Whelan}, {Lim}, {Earl}, {Starkman}, {Bradley}, {Shupe}, {Patil}, {Corrales}, {Brasseur}, {N{\"o}the}, {Donath}, {Tollerud}, {Morris}, {Ginsburg}, {Vaher}, {Weaver}, {Tocknell}, {Jamieson}, {van Kerkwijk}, {Robitaille}, {Merry}, {Bachetti}, {G{\"u}nther}, {Aldcroft}, {Alvarado-Montes}, {Archibald}, {B{\'o}di}, {Bapat}, {Barentsen}, {Baz{\'a}n}, {Biswas}, {Boquien}, {Burke}, {Cara}, {Cara}, {Conroy}, {Conseil}, {Craig}, {Cross}, {Cruz}, {D'Eugenio}, {Dencheva}, {Devillepoix}, {Dietrich}, {Eigenbrot}, {Erben}, {Ferreira}, {Foreman-Mackey}, {Fox}, {Freij}, {Garg}, {Geda}, {Glattly}, {Gondhalekar}, {Gordon}, {Grant}, {Greenfield}, {Groener}, {Guest}, {Gurovich}, {Handberg}, {Hart}, {Hatfield-Dodds}, {Homeier}, {Hosseinzadeh}, {Jenness}, {Jones}, {Joseph}, {Kalmbach}, {Karamehmetoglu}, {Ka{\l}uszy{\'n}ski}, {Kelley}, {Kern}, {Kerzendorf}, {Koch}, {Kulumani}, {Lee}, {Ly}, {Ma}, {MacBride}, {Maljaars}, {Muna}, {Murphy}, {Norman},
  {O'Steen}, {Oman}, {Pacifici}, {Pascual}, {Pascual-Granado}, {Patil}, {Perren}, {Pickering}, {Rastogi}, {Roulston}, {Ryan}, {Rykoff}, {Sabater}, {Sakurikar}, {Salgado}, {Sanghi}, {Saunders}, {Savchenko}, {Schwardt}, {Seifert-Eckert}, {Shih}, {Jain}, {Shukla}, {Sick}, {Simpson}, {Singanamalla}, {Singer}, {Singhal}, {Sinha}, {Sip{\H{o}}cz}, {Spitler}, {Stansby}, {Streicher}, {{\v{S}}umak}, {Swinbank}, {Taranu}, {Tewary}, {Tremblay}, {de Val-Borro}, {Van Kooten}, {Vasovi{\'c}}, {Verma}, {de Miranda Cardoso}, {Williams}, {Wilson}, {Winkel}, {Wood-Vasey}, {Xue}, {Yoachim}, {Zhang}, {Zonca}, \& {Astropy Project Contributors}}]{Astropy2022}
{Astropy Collaboration}, {Price-Whelan}, A.~M., {Lim}, P.~L., {et~al.} 2022, \apj, 935, 167

\bibitem[{{Belloni}(2010)}]{Belloni2010}
{Belloni}, T.~M. 2010, arXiv e-prints, arXiv:1007.5404

\bibitem[{{Berry} \& {Burnell}(2005)}]{Berry2005}
{Berry}, R. \& {Burnell}, J. 2005, The handbook of astronomical image processing, Vol.~2 (Richmond, Virginia: Willman-Bell)

\bibitem[{{Boeshaar}(1976)}]{Boeshaar76}
{Boeshaar}, P.~C. 1976, PhD thesis, The Ohio State University

\bibitem[{{Britt} {et~al.}(2012){Britt}, {Johnson}, \& {Hynes}}]{Britt2012}
{Britt}, C.~T., {Johnson}, C. C.~B., \& {Hynes}, R.~I. 2012, The Astronomer's Telegram, 4195

\bibitem[{{Cardelli} {et~al.}(1989){Cardelli}, {Clayton}, \& {Mathis}}]{Cardelli1989}
{Cardelli}, J.~A., {Clayton}, G.~C., \& {Mathis}, J.~S. 1989, \apj, 345, 245

\bibitem[{{Casares}(2015)}]{Casares2015}
{Casares}, J. 2015, \apj, 808, 80

\bibitem[{{Casares}(2016)}]{Casares2016}
{Casares}, J. 2016, \apj, 822, 99

\bibitem[{{Casares} {et~al.}(2022){Casares}, {Mu{\~n}oz-Darias}, {Torres}, {Mata S{\'a}nchez}, {Britt}, {Armas Padilla}, {{\'A}lvarez-Hern{\'a}ndez}, {C{\'u}neo}, {Gonz{\'a}lez Hern{\'a}ndez}, {Jim{\'e}nez-Ibarra}, {Jonker}, {Panizo-Espinar}, {S{\'a}nchez-Sierras}, \& {Yanes-Rizo}}]{Casares2022}
{Casares}, J., {Mu{\~n}oz-Darias}, T., {Torres}, M.~A.~P., {et~al.} 2022, \mnras, 516, 2023

\bibitem[{Casares {et~al.}(2012)Casares, Rodr\'iguez-Gil, Zurita, Corral-Santana, Mart\'inez-Pais, Corradi, Cornelisse, \& Charles}]{Casares2012}
Casares, J., Rodr\'iguez-Gil, P., Zurita, C., {et~al.} 2012, The Astronomer's Telegram, 4347

\bibitem[{Cenko \& Ofek(2012)}]{Cenko2012}
Cenko, S.~B. \& Ofek, E.~O. 2012, The Astronomer's Telegram, 4146

\bibitem[{{Cepa} {et~al.}(2003){Cepa}, {Aguiar-Gonzalez}, {Bland-Hawthorn}, {Castaneda}, {Cobos}, {Correa}, {Espejo}, {Fragoso-Lopez}, {Fuentes}, {Gigante}, {Gonzalez}, {Gonzalez-Escalera}, {Gonzalez-Serrano}, {Joven-Alvarez}, {Lopez-Ruiz}, {Militello}, {Cano}, {Perez}, {Perez}, {Rasilla}, {Sanchez}, \& {Tejada}}]{Cepa2003}
{Cepa}, J., {Aguiar-Gonzalez}, M., {Bland-Hawthorn}, J., {et~al.} 2003, in Society of Photo-Optical Instrumentation Engineers (SPIE) Conference Series, Vol. 4841, Instrument Design and Performance for Optical/Infrared Ground-based Telescopes, ed. M.~{Iye} \& A.~F.~M. {Moorwood}, 1739--1749

\bibitem[{{Chambers} {et~al.}(2016){Chambers}, {Magnier}, {Metcalfe}, {Flewelling}, {Huber}, {Waters}, {Denneau}, {Draper}, {Farrow}, {Finkbeiner}, {Holmberg}, {Koppenhoefer}, {Price}, {Rest}, {Saglia}, {Schlafly}, {Smartt}, {Sweeney}, {Wainscoat}, {Burgett}, {Chastel}, {Grav}, {Heasley}, {Hodapp}, {Jedicke}, {Kaiser}, {Kudritzki}, {Luppino}, {Lupton}, {Monet}, {Morgan}, {Onaka}, {Shiao}, {Stubbs}, {Tonry}, {White}, {Ba{\~n}ados}, {Bell}, {Bender}, {Bernard}, {Boegner}, {Boffi}, {Botticella}, {Calamida}, {Casertano}, {Chen}, {Chen}, {Cole}, {Deacon}, {Frenk}, {Fitzsimmons}, {Gezari}, {Gibbs}, {Goessl}, {Goggia}, {Gourgue}, {Goldman}, {Grant}, {Grebel}, {Hambly}, {Hasinger}, {Heavens}, {Heckman}, {Henderson}, {Henning}, {Holman}, {Hopp}, {Ip}, {Isani}, {Jackson}, {Keyes}, {Koekemoer}, {Kotak}, {Le}, {Liska}, {Long}, {Lucey}, {Liu}, {Martin}, {Masci}, {McLean}, {Mindel}, {Misra}, {Morganson}, {Murphy}, {Obaika}, {Narayan}, {Nieto-Santisteban}, {Norberg}, {Peacock}, {Pier}, {Postman}, {Primak}, {Rae}, {Rai},
  {Riess}, {Riffeser}, {Rix}, {R{\"o}ser}, {Russel}, {Rutz}, {Schilbach}, {Schultz}, {Scolnic}, {Strolger}, {Szalay}, {Seitz}, {Small}, {Smith}, {Soderblom}, {Taylor}, {Thomson}, {Taylor}, {Thakar}, {Thiel}, {Thilker}, {Unger}, {Urata}, {Valenti}, {Wagner}, {Walder}, {Walter}, {Watters}, {Werner}, {Wood-Vasey}, \& {Wyse}}]{Chambers2016}
{Chambers}, K.~C., {Magnier}, E.~A., {Metcalfe}, N., {et~al.} 2016, arXiv e-prints, arXiv:1612.05560

\bibitem[{Charles {et~al.}(2012)Charles, Cornelisse, \& Casares}]{Charles2012}
Charles, P., Cornelisse, R., \& Casares, J. 2012, The Astronomer's Telegram, 4210

\bibitem[{{Charles} \& {Coe}(2006)}]{Charles2006}
{Charles}, P.~A. \& {Coe}, M.~J. 2006, Compact Stellar X-ray Sources (Cambridge University Press), 215--265

\bibitem[{{Corral-Santana} {et~al.}(2016){Corral-Santana}, {Casares}, {Mu{\~n}oz-Darias}, {Bauer}, {Mart{\'{\i}}nez-Pais}, \& {Russell}}]{Corral-Santana2016}
{Corral-Santana}, J.~M., {Casares}, J., {Mu{\~n}oz-Darias}, T., {et~al.} 2016, \aap, 587, A61

\bibitem[{{Czesla} {et~al.}(2019){Czesla}, {Schr{\"o}ter}, {Schneider}, {Huber}, {Pfeifer}, {Andreasen}, \& {Zechmeister}}]{Czesla2019}
{Czesla}, S., {Schr{\"o}ter}, S., {Schneider}, C.~P., {et~al.} 2019, {PyA: Python astronomy-related packages}

\bibitem[{Degenaar {et~al.}(2014)Degenaar, Maitra, Cackett, Reynolds, Miller, Reis, King, Gültekin, Bailyn, Buxton, MacDonald, Fabian, Fox, \& Rykoff}]{Degenaar2014}
Degenaar, N., Maitra, D., Cackett, E.~M., {et~al.} 2014, \apj, 784, 122

\bibitem[{{Done} {et~al.}(2007){Done}, {Gierli{\'n}ski}, \& {Kubota}}]{Done2007}
{Done}, C., {Gierli{\'n}ski}, M., \& {Kubota}, A. 2007, \aapr, 15, 1

\bibitem[{{Fender} {et~al.}(2004){Fender}, {Belloni}, \& {Gallo}}]{Fender2004}
{Fender}, R.~P., {Belloni}, T.~M., \& {Gallo}, E. 2004, \mnras, 355, 1105

\bibitem[{{Gaia Collaboration} {et~al.}(2023){Gaia Collaboration}, {Vallenari}, {Brown}, {Prusti}, {de Bruijne}, {Arenou}, {Babusiaux}, {Biermann}, {Creevey}, {Ducourant}, {Evans}, {Eyer}, {Guerra}, {Hutton}, {Jordi}, {Klioner}, {Lammers}, {Lindegren}, {Luri}, {Mignard}, {Panem}, {Pourbaix}, {Randich}, {Sartoretti}, {Soubiran}, {Tanga}, {Walton}, {Bailer-Jones}, {Bastian}, {Drimmel}, {Jansen}, {Katz}, {Lattanzi}, {van Leeuwen}, {Bakker}, {Cacciari}, {Casta{\~n}eda}, {De Angeli}, {Fabricius}, {Fouesneau}, {Fr{\'e}mat}, {Galluccio}, {Guerrier}, {Heiter}, {Masana}, {Messineo}, {Mowlavi}, {Nicolas}, {Nienartowicz}, {Pailler}, {Panuzzo}, {Riclet}, {Roux}, {Seabroke}, {Sordo}, {Th{\'e}venin}, {Gracia-Abril}, {Portell}, {Teyssier}, {Altmann}, {Andrae}, {Audard}, {Bellas-Velidis}, {Benson}, {Berthier}, {Blomme}, {Burgess}, {Busonero}, {Busso}, {C{\'a}novas}, {Carry}, {Cellino}, {Cheek}, {Clementini}, {Damerdji}, {Davidson}, {de Teodoro}, {Nu{\~n}ez Campos}, {Delchambre}, {Dell'Oro}, {Esquej},
  {Fern{\'a}ndez-Hern{\'a}ndez}, {Fraile}, {Garabato}, {Garc{\'\i}a-Lario}, {Gosset}, {Haigron}, {Halbwachs}, {Hambly}, {Harrison}, {Hern{\'a}ndez}, {Hestroffer}, {Hodgkin}, {Holl}, {Jan{\ss}en}, {Jevardat de Fombelle}, {Jordan}, {Krone-Martins}, {Lanzafame}, {L{\"o}ffler}, {Marchal}, {Marrese}, {Moitinho}, {Muinonen}, {Osborne}, {Pancino}, {Pauwels}, {Recio-Blanco}, {Reyl{\'e}}, {Riello}, {Rimoldini}, {Roegiers}, {Rybizki}, {Sarro}, {Siopis}, {Smith}, {Sozzetti}, {Utrilla}, {van Leeuwen}, {Abbas}, {{\'A}brah{\'a}m}, {Abreu Aramburu}, {Aerts}, {Aguado}, {Ajaj}, {Aldea-Montero}, {Altavilla}, {{\'A}lvarez}, {Alves}, {Anders}, {Anderson}, {Anglada Varela}, {Antoja}, {Baines}, {Baker}, {Balaguer-N{\'u}{\~n}ez}, {Balbinot}, {Balog}, {Barache}, {Barbato}, {Barros}, {Barstow}, {Bartolom{\'e}}, {Bassilana}, {Bauchet}, {Becciani}, {Bellazzini}, {Berihuete}, {Bernet}, {Bertone}, {Bianchi}, {Binnenfeld}, {Blanco-Cuaresma}, {Blazere}, {Boch}, {Bombrun}, {Bossini}, {Bouquillon}, {Bragaglia}, {Bramante}, {Breedt},
  {Bressan}, {Brouillet}, {Brugaletta}, {Bucciarelli}, {Burlacu}, {Butkevich}, {Buzzi}, {Caffau}, {Cancelliere}, {Cantat-Gaudin}, {Carballo}, {Carlucci}, {Carnerero}, {Carrasco}, {Casamiquela}, {Castellani}, {Castro-Ginard}, {Chaoul}, {Charlot}, {Chemin}, {Chiaramida}, {Chiavassa}, {Chornay}, {Comoretto}, {Contursi}, {Cooper}, {Cornez}, {Cowell}, {Crifo}, {Cropper}, {Crosta}, {Crowley}, {Dafonte}, {Dapergolas}, {David}, {David}, {de Laverny}, {De Luise}, \& {De March}}]{GaiaDR3}
{Gaia Collaboration}, {Vallenari}, A., {Brown}, A.~G.~A., {et~al.} 2023, \aap, 674, A1

\bibitem[{{Green} {et~al.}(2019){Green}, {Schlafly}, {Zucker}, {Speagle}, \& {Finkbeiner}}]{Green2019}
{Green}, G.~M., {Schlafly}, E., {Zucker}, C., {Speagle}, J.~S., \& {Finkbeiner}, D. 2019, \apj, 887, 93

\bibitem[{{G{\"u}ver} \& {{\"O}zel}(2009)}]{Guver2009}
{G{\"u}ver}, T. \& {{\"O}zel}, F. 2009, \mnras, 400, 2050

\bibitem[{{Hamuy} {et~al.}(1994){Hamuy}, {Suntzeff}, {Heathcote}, {Walker}, {Gigoux}, \& {Phillips}}]{Hamuy1994}
{Hamuy}, M., {Suntzeff}, N.~B., {Heathcote}, S.~R., {et~al.} 1994, \pasp, 106, 566

\bibitem[{{Hamuy} {et~al.}(1992){Hamuy}, {Walker}, {Suntzeff}, {Gigoux}, {Heathcote}, \& {Phillips}}]{Hamuy1992}
{Hamuy}, M., {Walker}, A.~R., {Suntzeff}, N.~B., {et~al.} 1992, \pasp, 104, 533

\bibitem[{{Harvey} {et~al.}(1995){Harvey}, {Skillman}, {Patterson}, \& {Ringwald}}]{Harvey1995}
{Harvey}, D., {Skillman}, D.~R., {Patterson}, J., \& {Ringwald}, F.~A. 1995, \pasp, 107, 551

\bibitem[{{Haswell} {et~al.}(2002){Haswell}, {Hynes}, {King}, \& {Schenker}}]{Haswell2002}
{Haswell}, C.~A., {Hynes}, R.~I., {King}, A.~R., \& {Schenker}, K. 2002, \mnras, 332, 928

\bibitem[{{Haswell} {et~al.}(2001){Haswell}, {King}, {Murray}, \& {Charles}}]{Haswell2001}
{Haswell}, C.~A., {King}, A.~R., {Murray}, J.~R., \& {Charles}, P.~A. 2001, \mnras, 321, 475

\bibitem[{{Hosokawa} {et~al.}(2022){Hosokawa}, {Murata}, {Niwano}, {Ito}, {Takamatsu}, {Imai}, {Sato}, {Takaku}, {Noto}, {Yamaguchi}, {Yatsu}, \& {Kawai}}]{Hosokawa2022}
{Hosokawa}, R., {Murata}, K.~L., {Niwano}, M., {et~al.} 2022, The Astronomer's Telegram, 15226

\bibitem[{{Jim{\'e}nez-Ibarra} {et~al.}(2019){Jim{\'e}nez-Ibarra}, {Mu{\~n}oz-Darias}, {Armas Padilla}, {Russell}, {Casares}, {Torres}, {Mata S{\'a}nchez}, {Jonker}, \& {Lewis}}]{JimenezIbarra2019}
{Jim{\'e}nez-Ibarra}, F., {Mu{\~n}oz-Darias}, T., {Armas Padilla}, M., {et~al.} 2019, \mnras, 484, 2078

\bibitem[{{Kato}(2002)}]{Kato2002}
{Kato}, T. 2002, \pasj, 54, L11

\bibitem[{{Kato}(2015)}]{Kato2015}
{Kato}, T. 2015, \pasj, 67, 108

\bibitem[{Kimura {et~al.}(2012)Kimura, Tomida, Nakahira, Negoro, Ueno, Ishikawa, Mihara, Serino, Sugizaki, Yamamoto, Sugimoto, Matsuoka, Kawai, Morii, Usui, Ishikawa, Nakajima, Asada, Sakakibara, Serita, Yoshida, Tsunemi, Ueda, Hiroi, Shidatsu, Sato, Yamauchi, Nishimura, Hanayama, Yoshidome, Tsuboi, Higa, \& Yamaoka}]{Kimura2012}
Kimura, M., Tomida, H., Nakahira, S., {et~al.} 2012, The Astronomer's Telegram, 4198

\bibitem[{{Knigge} {et~al.}(2011){Knigge}, {Baraffe}, \& {Patterson}}]{Knigge2011}
{Knigge}, C., {Baraffe}, I., \& {Patterson}, J. 2011, \apjs, 194, 28

\bibitem[{{Kong}(2022)}]{Kong2022}
{Kong}, A.~K.~H. 2022, The Astronomer's Telegram, 15229

\bibitem[{{Kre{\l}owski} {et~al.}(2019){Kre{\l}owski}, {Galazutdinov}, {Godunova}, \& {Bondar}}]{Krelowski2019}
{Kre{\l}owski}, J., {Galazutdinov}, G., {Godunova}, V., \& {Bondar}, A. 2019, \actaa, 69, 159

\bibitem[{Krimm {et~al.}(2012)Krimm, Barthelmy, Baumgartner, Cummings, Fenimore, Gehrels, Markwardt, Palmer, Sakamoto, Skinner, Stamatikos, Tueller, \& Ukwatta}]{Krimm2012}
Krimm, H.~A., Barthelmy, S.~D., Baumgartner, W., {et~al.} 2012, The Astronomer's Telegram, 4139

\bibitem[{{Kuroda} {et~al.}(2010){Kuroda}, {Hanayama}, {Miyaji}, {Yanagisawa}, {Shimizu}, {Toda}, {Nagayama}, {Watanabe}, {Ali}, {Haroon}, {Essam}, {Ismail}, {Ismail}, {Kawabata}, {Yoshida}, {Ohta}, {Yatsu}, {Nakajima}, {Enomoto}, {Kawakami}, {Tokoyoda}, \& {Kawai}}]{Kuroda2010}
{Kuroda}, D., {Hanayama}, H., {Miyaji}, T., {et~al.} 2010, in The First Year of MAXI: Monitoring Variable X-ray Sources, 4

\bibitem[{{Lan} {et~al.}(2015){Lan}, {M{\'e}nard}, \& {Zhu}}]{Lan2015}
{Lan}, T.-W., {M{\'e}nard}, B., \& {Zhu}, G. 2015, \mnras, 452, 3629

\bibitem[{{Lasota}(2001)}]{Lasota2001}
{Lasota}, J.-P. 2001, \nar, 45, 449

\bibitem[{{Littlefair} {et~al.}(2006){Littlefair}, {Dhillon}, {Marsh}, \& {G{\"a}nsicke}}]{Littlefair2006}
{Littlefair}, S.~P., {Dhillon}, V.~S., {Marsh}, T.~R., \& {G{\"a}nsicke}, B.~T. 2006, \mnras, 371, 1435

\bibitem[{Lloyd {et~al.}(2012)Lloyd, Oksanen, Starr, Darlington, \& Pickard}]{Lloyd2012}
Lloyd, C., Oksanen, A., Starr, P., Darlington, G., \& Pickard, R. 2012, The Astronomer's Telegram, 4246

\bibitem[{{L{\'o}pez} {et~al.}(2019){L{\'o}pez}, {Jonker}, {Torres}, {Heida}, {Rau}, \& {Steeghs}}]{Lopez2019}
{L{\'o}pez}, K.~M., {Jonker}, P.~G., {Torres}, M.~A.~P., {et~al.} 2019, \mnras, 482, 2149

\bibitem[{{Marsh}(1990)}]{Marsh90}
{Marsh}, T.~R. 1990, \apj, 357, 621

\bibitem[{{McAllister} {et~al.}(2019){McAllister}, {Littlefair}, {Parsons}, {Dhillon}, {Marsh}, {G{\"a}nsicke}, {Breedt}, {Copperwheat}, {Green}, {Knigge}, {Sahman}, {Dyer}, {Kerry}, {Ashley}, {Irawati}, \& {Rattanasoon}}]{McAllister2019}
{McAllister}, M., {Littlefair}, S.~P., {Parsons}, S.~G., {et~al.} 2019, \mnras, 486, 5535

\bibitem[{{Mu{\~n}oz-Darias} {et~al.}(2013){Mu{\~n}oz-Darias}, {Coriat}, {Plant}, {Ponti}, {Fender}, \& {Dunn}}]{MunozDarias2013}
{Mu{\~n}oz-Darias}, T., {Coriat}, M., {Plant}, D.~S., {et~al.} 2013, \mnras, 432, 1330

\bibitem[{{Mu{\~n}oz-Darias} {et~al.}(2019){Mu{\~n}oz-Darias}, {Jim{\'e}nez-Ibarra}, {Panizo-Espinar}, {Casares}, {Mata S{\'a}nchez}, {Ponti}, {Fender}, {Buckley}, {Garnavich}, {Torres}, {Armas Padilla}, {Charles}, {Corral-Santana}, {Kajava}, {Kotze}, {Littlefield}, {S{\'a}nchez-Sierras}, {Steeghs}, \& {Thomas}}]{Munoz-Darias2019}
{Mu{\~n}oz-Darias}, T., {Jim{\'e}nez-Ibarra}, F., {Panizo-Espinar}, G., {et~al.} 2019, \apjl, 879, L4

\bibitem[{{Munari} {et~al.}(2008){Munari}, {Tomasella}, {Fiorucci}, {Bienaym{\'e}}, {Binney}, {Bland-Hawthorn}, {Boeche}, {Campbell}, {Freeman}, {Gibson}, {Gilmore}, {Grebel}, {Helmi}, {Navarro}, {Parker}, {Seabroke}, {Siebert}, {Siviero}, {Steinmetz}, {Watson}, {Williams}, {Wyse}, \& {Zwitter}}]{Munari2008}
{Munari}, U., {Tomasella}, L., {Fiorucci}, M., {et~al.} 2008, \aap, 488, 969

\bibitem[{Nakahira {et~al.}(2014)Nakahira, Negoro, Shidatsu, Ueda, Mihara, Sugizaki, Matsuoka, \& Onodera}]{Nakahira2014}
Nakahira, S., Negoro, H., Shidatsu, M., {et~al.} 2014, \pasj, 66, 84

\bibitem[{Nakahira {et~al.}(2012)Nakahira, Ueda, Takagi, Mihara, Sugizaki, Serino, Yamamoto, Sugimoto, Matsuoka, Ueno, Tomida, Ishikawa, Kawai, Morii, Usui, Ishikawa, Yoshii, Negoro, Nakajima, Asada, Sakakibara, Serita, Yoshida, Tsunemi, Kimura, Hiroi, Shidatsu, Sato, Tsuboi, Yamauchi, Nishimura, Hanayama, Yoshidome, \& Yamaoka}]{Nakahira2012}
Nakahira, S., Ueda, Y., Takagi, T., {et~al.} 2012, The Astronomer's Telegram, 4273

\bibitem[{{Nakata} {et~al.}(2013){Nakata}, {Ohshima}, {Kato}, {Nogami}, {Masi}, {de Miguel}, {Ulowetz}, {Littlefield}, {Goff}, {Krajci}, {Maehara}, {Stein}, {Sabo}, {Noguchi}, {Ono}, {Kawabata}, {Furukawa}, {Matsumoto}, {Ishibashi}, {Dubovsky}, {Kudzej}, {Dvorak}, {Hambsch}, {Pickard}, {Morelle}, {Muyllaert}, {Padovan}, \& {Henden}}]{Nakata2013}
{Nakata}, C., {Ohshima}, T., {Kato}, T., {et~al.} 2013, \pasj, 65, 117

\bibitem[{{Oke}(1990)}]{Oke1990}
{Oke}, J.~B. 1990, \aj, 99, 1621

\bibitem[{{Ponti} {et~al.}(2012){Ponti}, {Fender}, {Begelman}, {Dunn}, {Neilsen}, \& {Coriat}}]{Ponti2012}
{Ponti}, G., {Fender}, R.~P., {Begelman}, M.~C., {et~al.} 2012, \mnras, 422, L11

\bibitem[{{Ratti} {et~al.}(2012){Ratti}, {Steeghs}, {Jonker}, {Torres}, {Bassa}, \& {Verbunt}}]{rattietal12-1}
{Ratti}, E.~M., {Steeghs}, D.~T.~H., {Jonker}, P.~G., {et~al.} 2012, \mnras, 420, 75

\bibitem[{Rau {et~al.}(2012)Rau, Greiner, \& Schady}]{Rau2012}
Rau, A., Greiner, J., \& Schady, P. 2012, The Astronomer's Telegram, 4144

\bibitem[{Robitaille \& Bressert(2012)}]{Robitaille2012}
Robitaille, T. \& Bressert, E. 2012, {APLpy: Astronomical Plotting Library in Python}, Astrophysics Source Code Library, record ascl:1208.017, aSCL entry ascl:1208.017

\bibitem[{{Rodr{\'\i}guez-Gil} {et~al.}(2015){Rodr{\'\i}guez-Gil}, {Shahbaz}, {Marsh}, {G{\"a}nsicke}, {Steeghs}, {Long}, {Mart{\'\i}nez-Pais}, {Armas Padilla}, {Schwarz}, {Schreiber}, {Torres}, {Koester}, {Dhillon}, {Castellano}, \& {Rodr{\'\i}guez}}]{rodriguez-giletal15-1}
{Rodr{\'\i}guez-Gil}, P., {Shahbaz}, T., {Marsh}, T.~R., {et~al.} 2015, \mnras, 452, 146

\bibitem[{{Rodr{\'\i}guez-Gil} {et~al.}(2020){Rodr{\'\i}guez-Gil}, {Shahbaz}, {Torres}, {G{\"a}nsicke}, {Izquierdo}, {Toloza}, {{\'A}lvarez-Hern{\'a}ndez}, {Steeghs}, {van Spaandonk}, {Koester}, \& {Rodr{\'\i}guez}}]{rodriguez-giletal20-1}
{Rodr{\'\i}guez-Gil}, P., {Shahbaz}, T., {Torres}, M.~A.~P., {et~al.} 2020, \mnras, 494, 425

\bibitem[{{Russell} {et~al.}(2014){Russell}, {Soria}, {Motch}, {Pakull}, {Torres}, {Curran}, {Jonker}, \& {Miller-Jones}}]{Russell2014}
{Russell}, T.~D., {Soria}, R., {Motch}, C., {et~al.} 2014, \mnras, 439, 1381

\bibitem[{{Saikia} {et~al.}(2022){Saikia}, {Russell}, {Alabarta}, {Baglio}, {Bramich}, \& {Lewis}}]{Saikia2022}
{Saikia}, P., {Russell}, D.~M., {Alabarta}, K., {et~al.} 2022, The Astronomer's Telegram, 15303

\bibitem[{{Saikia} {et~al.}(2023{\natexlab{a}}){Saikia}, {Russell}, {Pirbhoy}, {Baglio}, {Bramich}, {Alabarta}, {Lewis}, \& {Charles}}]{Saikia2023a}
{Saikia}, P., {Russell}, D.~M., {Pirbhoy}, S.~F., {et~al.} 2023{\natexlab{a}}, \mnras, 524, 4543

\bibitem[{{Saikia} {et~al.}(2023{\natexlab{b}}){Saikia}, {Russell}, {Pirbhoy}, {Baglio}, {Bramich}, {Alabarta}, {Lewis}, \& {Charles}}]{Saikia2023}
{Saikia}, P., {Russell}, D.~M., {Pirbhoy}, S.~F., {et~al.} 2023{\natexlab{b}}, \apj, 949, 104

\bibitem[{{Schlafly} \& {Finkbeiner}(2011)}]{Schlafly2011}
{Schlafly}, E.~F. \& {Finkbeiner}, D.~P. 2011, \apjl, 737, 103

\bibitem[{{Schneider} \& {Young}(1980)}]{Schneider1980}
{Schneider}, D.~P. \& {Young}, P. 1980, \apj, 238, 946

\bibitem[{{Schwarzenberg-Czerny}(1996)}]{Schwarzenberg-Czerny1996}
{Schwarzenberg-Czerny}, A. 1996, \apjl, 460, L107

\bibitem[{{Smith} \& {Dhillon}(1998)}]{Smith+Dhillon1998}
{Smith}, D.~A. \& {Dhillon}, V.~S. 1998, \mnras, 301, 767

\bibitem[{{Stellingwerf}(1978)}]{Stellingwerf1978}
{Stellingwerf}, R.~F. 1978, \apj, 224, 953

\bibitem[{{Stetson}(1987)}]{Stetson1987}
{Stetson}, P.~B. 1987, \pasp, 99, 191

\bibitem[{{Thomas} {et~al.}(2022){Thomas}, {Charles}, {Buckley}, {Kotze}, {Lasota}, {Potter}, {Steiner}, \& {Paice}}]{Thomas2022}
{Thomas}, J.~K., {Charles}, P.~A., {Buckley}, D. A.~H., {et~al.} 2022, \mnras, 509, 1062

\bibitem[{{Tominaga} {et~al.}(2022){Tominaga}, {Nakahira}, {Negoro}, {Serino}, {Sugita}, {Nakajima}, {Kobayashi}, {Asakura}, {Seino}, {Mihara}, {Tamagawa}, {Li}, {Matsuoka}, {Sakamoto}, {Komachi}, {Hiramatsu}, {Yoshida}, {Tsuboi}, {Iwakiri}, {Kawai}, {Okamoto}, {Kitakoga}, {Kohara}, {Shidatsu}, {Iwasaki}, {Kawai}, {Niwano}, {Hosokawa}, {Imai}, {Ito}, {Takamatsu}, {Ueno}, {Tomida}, {Ishikawa}, {Nagatsuka}, {Kurihara}, {Ueda}, {Yamada}, {Ogawa}, {Setoguchi}, {Yoshitake}, {Goto}, {Uematsu}, {Inaba}, {Tsunemi}, {Yamauchi}, {Nonaka}, {Sato}, {Hatsuda}, {Fukuoka}, {Kawamuro}, {Yamaoka}, {Kawakubo}, \& {Sugizaki}}]{Tominaga2022}
{Tominaga}, M., {Nakahira}, S., {Negoro}, H., {et~al.} 2022, The Astronomer's Telegram, 15214

\bibitem[{{Tonry} {et~al.}(2012){Tonry}, {Stubbs}, {Lykke}, {Doherty}, {Shivvers}, {Burke}, {Chambers}, {Flewelling}, {Magnier}, {Morgan}, {Price}, {Peters}, {Roth}, {Smith}, \& {Weinberg}}]{Tonry2012}
{Tonry}, J.~L., {Stubbs}, C.~W., {Lykke}, K.~R., {et~al.} 2012, \apj, 750, 99

\bibitem[{{Torres} {et~al.}(2021){Torres}, {Jonker}, {Casares}, {Miller-Jones}, \& {Steeghs}}]{Torres2021}
{Torres}, M.~A.~P., {Jonker}, P.~G., {Casares}, J., {Miller-Jones}, J.~C.~A., \& {Steeghs}, D. 2021, \mnras, 501, 2174

\bibitem[{{Trelawny}(2013)}]{Trelawny2013}
{Trelawny}, D.~T. 2013, Master's thesis, California State University, Fresno, USA

\bibitem[{{Truss} {et~al.}(2002){Truss}, {Wynn}, {Murray}, \& {King}}]{Truss2002}
{Truss}, M.~R., {Wynn}, G.~A., {Murray}, J.~R., \& {King}, A.~R. 2002, \mnras, 337, 1329

\bibitem[{{Uemura} {et~al.}(2000){Uemura}, {Kato}, {Matsumoto}, {Yamaoka}, {Takamizawa}, {Sano}, {Haseda}, {Cook}, {Buczynski}, \& {Masi}}]{Uemura2000}
{Uemura}, M., {Kato}, T., {Matsumoto}, K., {et~al.} 2000, \pasj, 52, L15

\bibitem[{{Uemura} {et~al.}(2012){Uemura}, {Kato}, {Ohshima}, \& {Maehara}}]{Uemura2012}
{Uemura}, M., {Kato}, T., {Ohshima}, T., \& {Maehara}, H. 2012, \pasj, 64, 92

\bibitem[{Usui {et~al.}(2012)Usui, Nakahira, Tomida, Negoro, Morii, Kawai, Ishikawa, Ueno, Ishikawa, Mihara, Sugizaki, Serino, Yamamoto, Matsuoka, Yoshida, Tsunemi, Kimura, Nakajima, Asada, Sakakibara, Serita, Ueda, Hiroi, Shidatsu, Sato, Tsuboi, Yamauchi, Nishimura, Hanayama, Yoshidome, \& Yamaoka}]{Usui2012}
Usui, R., Nakahira, S., Tomida, H., {et~al.} 2012, The Astronomer's Telegram, 4140

\bibitem[{{VanderPlas}(2018)}]{VanderPlas2018}
{VanderPlas}, J.~T. 2018, \apjs, 236, 16

\bibitem[{{Verro} {et~al.}(2022){Verro}, {Trager}, {Peletier}, {Lan{\c{c}}on}, {Gonneau}, {Vazdekis}, {Prugniel}, {Chen}, {Coelho}, {S{\'a}nchez-Bl{\'a}zquez}, {Martins}, {Arentsen}, {Lyubenova}, {Falc{\'o}n-Barroso}, \& {Dries}}]{Verro2022}
{Verro}, K., {Trager}, S.~C., {Peletier}, R.~F., {et~al.} 2022, \aap, 660, A34

\bibitem[{Virtanen {et~al.}(2020)Virtanen, Gommers, Oliphant, Haberland, Reddy, Cournapeau, Burovski, Peterson, Weckesser, Bright, {van der Walt}, Brett, Wilson, Millman, Mayorov, Nelson, Jones, Kern, Larson, Carey, Polat, Feng, Moore, {VanderPlas}, Laxalde, Perktold, Cimrman, Henriksen, Quintero, Harris, Archibald, Ribeiro, Pedregosa, {van Mulbregt}, \& {SciPy 1.0 Contributors}}]{Virtanen2020}
Virtanen, P., Gommers, R., Oliphant, T.~E., {et~al.} 2020, Nature Methods, 17, 261

\bibitem[{{Wade} \& {Horne}(1988)}]{Wade+Horne88}
{Wade}, R.~A. \& {Horne}, K. 1988, \apj, 324, 411

\bibitem[{{Wenger} {et~al.}(2000){Wenger}, {Ochsenbein}, {Egret}, {Dubois}, {Bonnarel}, {Borde}, {Genova}, {Jasniewicz}, {Lalo{\"e}}, {Lesteven}, \& {Monier}}]{Wenger2000}
{Wenger}, M., {Ochsenbein}, F., {Egret}, D., {et~al.} 2000, \aaps, 143, 9

\bibitem[{{Williams} {et~al.}(2022){Williams}, {Motta}, {Rhodes}, {Fender}, {Bahramian}, {Green}, {Titterington}, \& {Sivakoff}}]{Williams2022}
{Williams}, D., {Motta}, S., {Rhodes}, L., {et~al.} 2022, The Astronomer's Telegram, 15219

\bibitem[{{Yanes-Rizo} {et~al.}(2025){Yanes-Rizo}, {Torres}, {Casares}, {Jonker}, {S{\'a}nchez-Sierras}, {Mu{\~n}oz-Darias}, \& {Armas Padilla}}]{YanesRizo2025}
{Yanes-Rizo}, I.~V., {Torres}, M.~A.~P., {Casares}, J., {et~al.} 2025, \aap, 694, A119

\bibitem[{{Young} \& {Schneider}(1981)}]{Young+Schneider81}
{Young}, P. \& {Schneider}, D.~P. 1981, \apj, 247, 960

\bibitem[{{Zechmeister} \& {K{\"u}rster}(2009)}]{Zechmeister2009}
{Zechmeister}, M. \& {K{\"u}rster}, M. 2009, \aap, 496, 577

\bibitem[{{Zhang} {et~al.}(2019){Zhang}, {Bernardini}, {Russell}, {Gelfand}, {Lasota}, {Qasim}, {AlMannaei}, {Koljonen}, {Shaw}, {Lewis}, {Tomsick}, {Plotkin}, {Miller-Jones}, {Maitra}, {Homan}, {Charles}, {Kobel}, {Perez}, \& {Doran}}]{Zhang2019}
{Zhang}, G.~B., {Bernardini}, F., {Russell}, D.~M., {et~al.} 2019, \apj, 876, 5

\bibitem[{{Zurita} {et~al.}(2002){Zurita}, {Casares}, {Shahbaz}, {Wagner}, {Foltz}, {Rodr{\'{\i}}guez-Gil}, {Hynes}, {Charles}, {Ryan}, {Schwarz}, \& {Starrfield}}]{Zurita2002}
{Zurita}, C., {Casares}, J., {Shahbaz}, T., {et~al.} 2002, \mnras, 333, 791

\end{thebibliography}

\label{LastPage}
\end{document}